\documentclass[12pt]{article}

\usepackage{natbib}
\usepackage[utf8]{inputenc} % allow utf-8 input
\usepackage[T1]{fontenc}    % use 8-bit T1 fonts
\usepackage{hyperref}   
\hypersetup{
	colorlinks=true,       % false: boxed links; true: colored links
	linkcolor=blue,        % color of internal links
	citecolor=blue,        % color of links to bibliography
	filecolor=magenta,     % color of file links
	urlcolor=blue
}% hyperlinks
\usepackage{url}            % simple URL typesetting
\usepackage{booktabs}       % professional-quality tables
\usepackage{amsfonts}       % blackboard math symbols
\usepackage{nicefrac}       % compact symbols for 1/2, etc.
\usepackage{microtype}      % microtypography
\usepackage{xcolor}         % colors
\usepackage{smile}
\usepackage{algorithm}
\usepackage{algorithmic}
\usepackage[]{graphicx,float,latexsym,times}
\usepackage{amsfonts,amstext,amsmath,amssymb,amsthm}
\usepackage{wrapfig}
\usepackage{lipsum}
\usepackage{ragged2e}
\usepackage{geometry}
\usepackage{setspace}   
\usepackage[symbol]{footmisc}
\renewcommand*{\thefootnote}{\fnsymbol{footnote}}

\usepackage[title,toc]{appendix}
\usepackage{minitoc}
\noptcrule
\title{Granger Causal Chain Discovery for Sepsis-Associated Derangements via Continuous-Time Hawkes Processes}

\author{Song Wei$^\mathrm{a}$\footnote{Correspondence to: Song Wei <song.wei@gatech.edu>.} , \quad Yao Xie$^\mathrm{a}$, \quad Christopher S. Josef$^\mathrm{b}$, \quad Rishikesan Kamaleswaran$^\mathrm{c,d}$\\
\\
  \small{$^\mathrm{a}$School of Industrial and Systems Engineering, Georgia Institute of Technology.} \\
  \small{$^\mathrm{b}$Department of Surgery, Emory University School of Medicine.}\\
  \small{$^\mathrm{c}$Department of Biomedical Informatics, Emory University School of Medicine.}\\
  \small{$^\mathrm{d}$Department of Biomedical Engineering, Georgia Institute of Technology.}
}
\date{\vspace{-20pt}}

\begin{document}

\maketitle

\begin{abstract}
Modern health care systems are conducting continuous, automated surveillance of the electronic medical record (EMR) to identify adverse events with increasing frequency; however, many events such as sepsis do not have elucidated prodromes (i.e., event chains) that can be used to identify and intercept the adverse event early in its course. Clinically relevant and interpretable results require a framework that can (i) infer temporal interactions across multiple patient features found in EMR data (e.g., Labs, vital signs, etc.) and (ii) identify patterns that precede and are specific to an impending adverse event (e.g., sepsis). In this work, we propose a linear multivariate Hawkes process model, coupled with ReLU link function, to recover a Granger Causal (GC) graph with both exciting and inhibiting effects. We develop a scalable two-phase gradient-based method to obtain a maximum surrogate-likelihood estimator, which is shown to be effective via extensive numerical simulation. Our method is subsequently extended to a data set of patients admitted to Grady hospital system in Atlanta, GA, USA, where the estimated GC graph identifies several highly interpretable GC chains that precede sepsis. The code is available at \url{https://github.com/SongWei-GT/two-phase-MHP}.
\end{abstract}


\doparttoc % Tell to minitoc to generate a toc for the parts
\faketableofcontents % Run a fake tableofcontents command for the partocs

\part{} % Start the document part
%\parttoc % Insert the document TOC

\renewcommand*{\thefootnote}{\arabic{footnote}}

\vspace{-.45in}
\section{Introduction}

Continuous, automated surveillance systems that use machine learning models to identify adverse patient events are being incorporated into healthcare environments with increasing frequency. One of the most notable adverse events is sepsis, a life-threatening medical condition contributing to one in five deaths globally \citep{world2020global} and stands as one of the most important cases for automated in-hospital surveillance. Sepsis is formally defined as life-threatening organ dysfunction caused by a dysregulated host response to infection \citep{singer2016third}. Delays in recognizing sepsis and initiating appropriate treatment can adversely impact patient outcomes. In a recent study of adult sepsis patients, each hour of delayed treatment was associated with higher risk-adjusted in-hospital mortality (odds ratio, 1.04 per hour) \citep{seymourTimeTreatmentMortality2017}. It logically follows that early recognition of the physiologic aberrations preceding sepsis would afford clinicians more time to intervene and may contribute to improving outcomes and reducing costs.
Many machine learning methods have been developed to predict the onset of sepsis, utilizing data from the electronic medical record (EMR) \citep{fleuren2020machine,reyna2019early,shashikumarDeepAISEInterpretableRecurrent2021}.
While many approaches can be designed to provide an alert preceding an event, most are not designed to discover and report the causal chains that preceded an adverse event. Developing and reporting a causal chain of events not only serves as a foundation for prognosticating adverse event occurrence, but more importantly it reveals the pathways of deterioration which may afford clinicians the additional context to corroborate or modify existing treatment modalities in a way that is superior to a simple alarm.

\begin{figure*}[!htp]
%%%%\vspace{-0.1in}
\centerline{
\includegraphics[width = .49\textwidth]{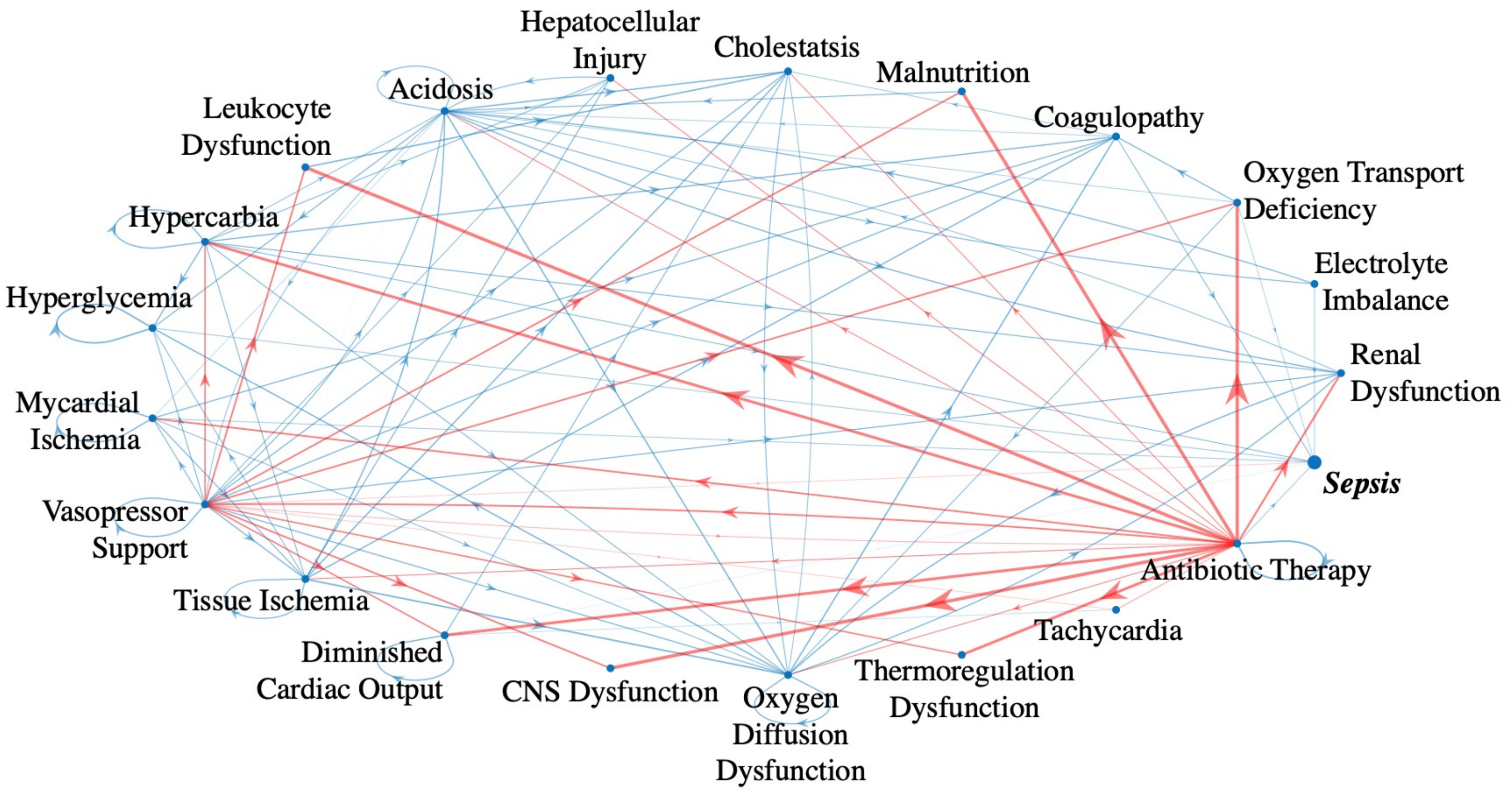}

\hspace{0.1in}

\includegraphics[width = .49\textwidth]{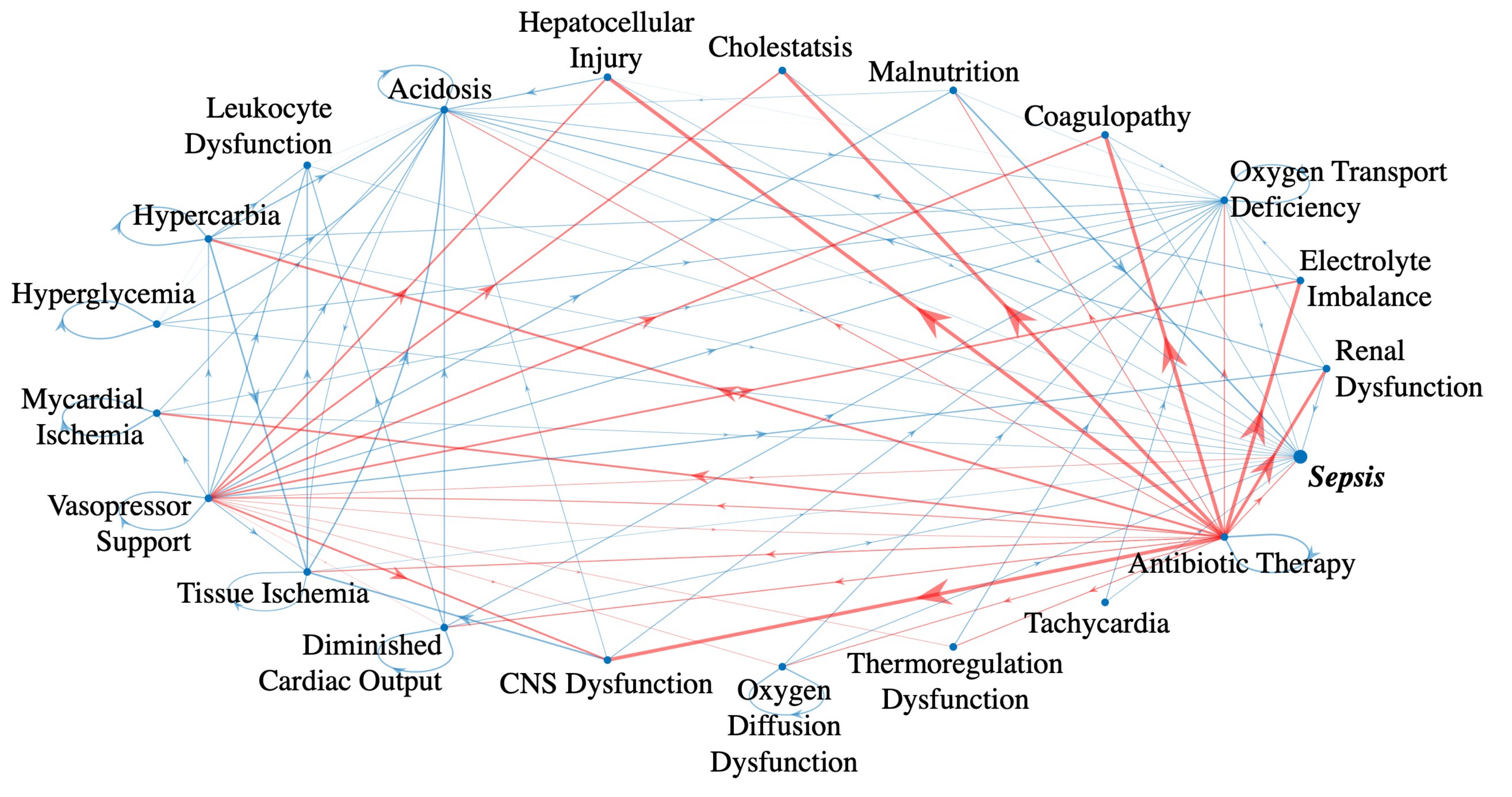}
}
%\vspace{-0.15in}
\caption{GC graphs over SADs for Sepsis-3 cohort (left) and full patient cohort (right). The width of the directed edge is proportional to the exciting (blue) or inhibiting (red) effect magnitude. We can observe that our proposed method can output highly interpretable GC graphs; for example, the observation that Antibiotic Therapy inhibits most of the SADs agrees with the well-known physiologic relationship.}
\label{fig:graph}
%\vspace{-0.15in}
\end{figure*}

Recently, Hawkes processes \citep{hawkes1971point,hawkes1971spectra,hawkes1974cluster}, which model self- and mutual- exciting patterns among continuous-time events, have drawn much attention in the field of health analytics \citep{meyer2014power,choi2015constructing,bao2017hawkes,schoenberg2019recursive,wei2021inferringb}.
The linear multivariate Hawkes process (MHP) seems highly relevant to our problem since 
(i) 
the support of the excitation matrix enjoys a natural interpretation as a Granger Causal (GC) graph \citep{xu2016learning},
(ii) 
given its interpretation as a clustering process \citep{hawkes1974cluster}, we can infer the commonly observed chain pattern that precedes sepsis from the estimated GC graph,
and 
(iii) with proper domain expertise, simple methods, such as (generalized) linear model, are proven effective in outputting highly explainable results \citep{choi2015constructing,wei2021inferringb}.

However, there are two major challenges preventing us from applying naive linear MHP to recover the GC graph. 
First and foremost, linear MHP itself fails to model inhibiting effects (e.g., proper medication will inhibit the occurrence of a certain disease), since ``negative triggering effects'' could lead to a negative conditional intensity and thus intractable likelihood. 
Second, the well-established expectation–maximization (EM) stochastic declustering algorithm \citep{zhuang2002stochastic,fox2016spatially} sufferers from scalability issue and cannot be applied to EMR data with thousands of patients' trajectories. 
Recently, \citet{bonnet2021maximum,bonnet2022inference} proposed a linear MHP coupled with ReLU link function $g(x) = x^+ := \max\{0,x\}$ to handle the potential inhibiting effects. To evaluate and maximize the likelihood, they calculated the ``re-start time'', at which the conditional intensity becomes nonzero. However, such a calculation has quadratic complexity, making it unscalable. Scalable methods to infer GC graph with both exciting and inhibiting effects for linear MHP are still largely missing.

% \begin{figure}[!htp]
% %%%%\vspace{-0.1in}
% \centerline{
% \includegraphics[width = 0.5\textwidth]{tex files/DAG_Hawkes/realdata/sep_graph.pdf}  }
% %%%%\vspace{-0.1in}
% \caption{Granger Causal graph over endogenous Sepsis Associated Derangement (SAD) indicators obtained via Discrete Hawkes Network with forward feature selection.}
% \label{fig:sep_graph} 
% %%%%\vspace{-0.1in}
% \end{figure}

%%%%\vspace{0.05in}

%\textbf{Contribution.}
In this paper, we adopt the ReLU link function in linear MHP to recover a Granger Causal graph with both exciting and inhibiting effects. 
We propose a maximum surrogate-likelihood formulation to tackle the scalability issue caused by the re-start time calculation \citep{bonnet2022inference}.
Furthermore, we develop a two-phase gradient-based method to solve the optimization problem, and we observe improved empirical performance through extensive numerical simulation.
Most importantly, our method can output graphs (i.e., Figure~\ref{fig:graph}) that afford clinicians a simple mechanism for interpreting both promoting and inhibitory causal relationship amongst the data --- Networks are exceptionally important for syndromic (i.e. a constellation of different physiologic derangements can be manifested) conditions like sepsis. These graphs can be used to differentiate cohorts and to identify important, intra-cohort relationships. For clinicians the utility of these graphs is two-fold: they can be used to (i) quantify a patient's risk of developing subsequent physiologic derangements in the future and (ii) discover new relationships. The estimated GC graphs here are highly interpretable and can be used to create or augment surveillance systems for high-risk patients. Here, we demonstrate the effectiveness of our approach in learning a Granger Causal graph for Sepsis Associated Derangements (SADs), but it can be generalized to other applications with similar requirements.

\paragraph{Related work.}
Granger Causality is well-studied in time series literature via the vector autoregressive (VAR) model; see \citet{shojaie2021granger} for a recent survey. 
%Recent advancements mainly focus on non-linear dynamics \citep{sindhwani2012scalable,tank2018neural}
%and tackling high-dimensionality via regularization \citep{bolstad2011causal,basu2015network,basu2019low}; for a comprehensive survey of the recent development of Granger Causality in the context of time series, we refer readers to \citet{shojaie2021granger}.
VAR models and MHP models share many similarities and some have recently recognized that the self- and mutual-excitation matrix in the Hawkes process model can be interpreted as Granger Causal graph in a similar way. 
The study of GC under the context of MHP can be traced back to \citet{kim2011granger}. Recent development includes leveraging the alternating direction method of multipliers to infer the low-rank structure in mutual excitation matrix \citep{zhou2013learning}, applying EM algorithm with various constraints \citep{xu2016learning,chen2022learning,ide2021cardinality} and using powerful neural networks \citep{zhang2020cause} to infer the GC graph.

Even outside the context of Granger Causality, the Hawkes process itself has drawn much attention recently --- there have been many (semi-)parametric Hawkes process models by considering different types of triggering kernel function, such as probability weighted kernel estimation with adaptive bandwidth \citep{zhuang2002stochastic}, probability weighted histogram estimation \citep{marsan2008extending} and with inhomogeneous spatial background rate \citep{fox2016spatially} and so on. In addition, there are also many non-parametric methods, e.g., the Neural Hawkes process \citep{mei2017neural} and the Transformer Hawkes process \citep{zuo2020transformer}.

Despite those advancements in semi- and non-parametric Hawkes process models, \citet{choi2015constructing,wei2021inferringb} showed that simple linear models can output meaningful results in practice. However, the state-of-the-art method is the stochastic declustering algorithm, which is based on the EM algorithm and is thus highly unscalable. This scalability issue makes it a less desirable option when we handle EMR data.
Recently, there are attempts to explore the powerful yet simple gradient-based method to infer the problem parameters; notable contributions include  \citet{wang2020uncertainty,cartea2021gradient}.
In particular, we want to mention that using the ReLU link to allow potential inhibiting effect in linear MHP was recently proposed by \citet{bonnet2021maximum,bonnet2022inference} and relatively novel in literature --- there have not been many methods tailored to this particular parameterization, and thus we only numerically compare our method with this re-start time method as well as some naive gradient-based methods.

Another closely related topic is causal discovery, which has drawn much attention in the past few decades. The state-of-the-art constraint-based algorithms include PC and Fast Causal Inference (FCI)  \citep{spirtes2000constructing}. Both algorithms can output the underlying true graph structure in the large sample limit. However, PC cannot deal with unobserved confounding whereas FCI is capable of dealing with confounders. However, since those algorithms rely on conditional independence tests to eliminate edges from the complete graph, they are not scalable when the number of nodes becomes large. Existing work to handle this includes a fast and memory-efficient PC algorithm using the parallel computing technique \citep{le2016fast}. Moreover, there is a continuous optimization-based approach to infer the underlying directed acyclic graph (DAG) structure, e.g., \citet{zheng2018dags}, which alleviates the aforementioned scalability issue. In addition, for time series data, existing causal discovery algorithms need to adapt to the potential temporal dependence. The most well-known method would be using AR time series to infer the Granger Causality, and later on, \citet{xu2016learning} extend GC to the context of the point process. However, the GC framework typically relies on the ``no unobserved confounding'' assumption. Examples to handle this include the FCI algorithm for time series to handle confounders \citep{entner2010causal}. It remains an open problem how to apply PC and FCI to point process data. For a complete survey on recent developments in causal inference, we refer readers to \citet{glymour2019review}.
%Our model can be viewed as a special case of Neural Hawkes process with no hidden layer. This is actually considered in the neural Hawkes process paper [7], even though they used softplus instead of ReLU as the activation function. 

%%%\vspace{-0.1in}
\section{Background}
%%%%\vspace{-0.05in}

\subsection{Multivariate Hawkes Process} 
Consider $d$ types events modeled by a counting process 
$N = (N^1, $ $\dots,N^d),$
where each process $N^i = \{N^i_t: t\in [0,T]\}$ itself is a counting measure on time horizon $T$ and records the number of type-$i$ events before time $t$. Such a process is called a linear MHP if the conditional intensity of $i$-th process $(i=1, \cdots, d)$ is defined as:
$$
\lambda_{i}(t)=\mu_{i}+\sum_{j=1}^d \int_{0}^t \varphi_{i,j}\left(s\right)d N_{t-s}^j,
$$
where $\mu_i$ is the exogenous background intensity for type-$i$ event and independent of the history, and kernel function $\varphi_{i, j}(\cdot)$ captures the impact from historical type-$j$ event to subsequent type-$i$ event.

Here, we adopt a very common and popular exponential kernel function $\varphi_{i, {j}}(t) = \alpha_{i,j}\exp\{-\beta t\}.$ The parameter $\alpha_{i,j}$ represents the magnitude of the impact from type-$j$ event to type-$i$ event and $\beta$ characterizes the rate of decay of that impact.
Most importantly, unlike the classic model, we consider both exciting and inhibiting effects by allowing negative magnitude parameters $\alpha_{i,j}$'s. However, this could lead to negative intensity, which contradicts the understanding of conditional intensity as the instantaneous probability of event occurrence. Following \citet{bonnet2021maximum}, we apply the {\it ReLU link function} $(\cdot)^+ = \max\{0,\cdot\}$ to the linear conditional intensity to fix this issue and get
\begin{equation}\label{eq:condi_intensity_def}
    \lambda_{i}(t)=\bigg(\mu_{i}+\sum_{j=1}^d \int_{0}^t \alpha_{i,j}e^{-\beta s}d N_{t-s}^j\bigg)^+.
\end{equation}
We denote the background intensity vector as $\mu = (\mu_1,\dots,\mu_d)^T$ and the self and mutual excitation/inhibition matrix as $A = (\alpha_{i,j}) \in \mathbb{R}^{d \times d}$. 
We will show the support of matrix $A$ can be interpreted as a Granger Causal graph.

%%%%\vspace{0.05in}

\subsection{Granger Causality}
In the seminal paper, \citet{eichler2017graphical} showed that the Granger Causal structure of the MHP is fully encoded in matrix $A$:
% The notion of Granger Causality was introduced to multivariate Hawkes process by \citet{eichler2017graphical,xu2016learning}. 
% Under our multivairate linear Hawkes process (with exponential decay kernel) model, the Granger Causality can be formally defined as follows:
\begin{proposition}[\citet{eichler2017graphical}]
Let $N = (N^1,\dots,N^d)$ be a $d$-dimensional multivariate Hawkes process with conditional intensity defined in \eqref{eq:condi_intensity_def}, then $N^j$ does NOT Granger-cause $N^i$ if and only if $\alpha_{i,j} = 0$.
\end{proposition}

%%%%\vspace{-0.05in}

We need to remark that inferring Granger Causality needs ``all the information in the universe'' 
%and hence we can only learn Granger non-causality and {\it prima facie causality} given partially observed data 
\citep{granger1969investigating,granger1980testing,granger1988some}. In the graph induced by the matrix $A = (\alpha_{i,j})$, the absence of an edge means Granger non-causality whereas only when there is no unobserved confounding can the presence of an edge in $A$ imply Granger causality. Here, we assume there is {\it no unobserved confounding} and we will take this matrix $A$ as the Granger Causal graph.
% , assuming that the observed information is ``enough'' to infer Granger Causality.

%By the above proposition, we can see the support of the excitation/inhibition matrix $A = (\alpha_{i,j})$ can be naturally taken as a Granger Causal graph. 
% In addition, this excitation/inhibition matrix $A$ can be understood as a directed information graph (DIG) \citep{etesami2016learning}, which is a generalized causal notion of Granger Causality. To be precise, in DIG, we determine the causality by comparing two conditional probabilities in KL-divergence sense: one is the conditional probability of $N^i_{t+dt}$ given full history, and the other one is the conditional probability of $N^i_{t+dt}$ given full history except that of type-$j$ event. Last but not least, both Granger Causal graph and DIG are equivalent to minimal generative model graphs \citep{quinn2011equivalence} and therefore can be used for causal inference in the same manner Bayesian networks are used for correlative statistical inference.

%%%\vspace{-0.1in}
\section{Estimation}\label{sec:method}
%%%%\vspace{-0.05in}

Consider the following continuous-time event data over a time horizon $T > 0$:
$$(u_1, t_1),\dots,(u_N, t_N),$$ 
where $0 \leq t_1 < \dots < t_N \leq T$ denote the exact occurrence times of the events and $u_n \in \{1,\dots,d\}$ represents the type of the $n$-th event.
The conditional intensity function of type-$i$ event at time $0 \leq t \leq T$ is as follows:
$$
\lambda_{i}(t)=\bigg(\mu_{i}+\sum_{j: t_{j}<t} \alpha_{i, u_{j}} e^{-\beta\left(t-t_{j}\right)}\bigg)^+.
$$ 
Typically, we use the Maximum likelihood estimation (MLE) to
learn model parameters, where the {\it true log likelihood} is:
\begin{equation}\label{eq:true_lik}
    \ell(\mu,A;\beta) = \sum_{i=1}^{d} \left( \int_{0}^{T} \log \lambda_{i}(t) d N_{t}^{i}-\int_{0}^{T} \lambda_{i}(t) d t \right).
\end{equation}

%%%\vspace{-0.1in}

\subsection{Existing method}

%%%\vspace{-0.05in}

In \eqref{eq:true_lik}, the first term reduces to a summation over the log-intensities on event occurrence times $\sum_{n = 1}^N \log \lambda_{u_n}(t_n)$, which will be well-defined since the conditional intensity at the event occurrence time will be positive. To be precise, the feasible region is 
\begin{equation}\label{eq:feasible_region}
    \Theta = \{(\mu,A): \tilde \lambda_{u_n}(t_n) > 0, \ n=1,\dots,N\},
\end{equation}
where the {\it surrogate conditional intensity} is defined as:
\begin{equation}\label{eq:surrogate_intensity}
    \tilde \lambda_{i}(t) = \mu_{i}+\sum_{j: t_{j}<t} \alpha_{i, u_{j}} e^{-\beta\left(t-t_{j}\right)}.
\end{equation}
After each event occurrence, due to the potential inhibiting effect, there could be an event with negative surrogate intensity; the ReLU link enforces such negative value to be zero and ensures that $\lambda_{i}(t) = (\tilde \lambda_{i}(t))^+$ is still a valid intensity. 
Nevertheless, it still takes some time for the process to ``re-start'', and we will call the time when the surrogate intensity increases to zero again as the ``re-start time''; see a graphical illustration in Figure~\ref{fig:restart} in the appendix.  %Appendix~\ref{appendix:benchmark}
To be precise, after the occurrence of $n$-th event $(u_n,t_n)$, the $n$-th re-start time for $i$-th process is as follows \citep{bonnet2021maximum,bonnet2022inference}:
\begin{equation}\label{eq:re_start_time}
    T_{(n,u_n)}^{(i)} = \min\left\{t_{n+1}, \ \arg \min_{t: \ t > t_n} \tilde \lambda_{i}(t) \geq 0\right\}.
\end{equation}
Then, for $(\mu,A) \in \Theta$, we can re-write \eqref{eq:true_lik} into:
\begin{align}
    \ell(\mu,A ;\beta)   =  \sum_{n = 1}^N \log  \tilde \lambda_{u_n}(t_n)  -  \sum_{i=1}^{d}\left(\int_{0}^{t_1}  + \sum_{n=1}^{N-1} \int_{T_{(n,u_n)}^{(i)}}^{t_{n+1}} +  \int_{T_{(N,u_N)}^{(i)}}^{T}\right) \tilde \lambda_{i}(t) d t. \label{eq:re_start_lik}
\end{align}

%%%\vspace{-0.1in}

Now, we remove the non-differentiable ReLU link in the log likelihood and the objective becomes differentiable. Despite its complicated form, the log likelihood objective can be calculated in closed-from due to the analytical expression of the re-start times \citep{bonnet2022inference}. Thus, we can leverage the powerful stochastic gradient descent (SGD) method to numerically solve the MLE.

%%%\vspace{-0.1in}

\subsection{Proposed gradient-based method}

\subsubsection{Empirical challenge}

\begin{figure}[!htp]
%%%\vspace{-0.05in}
\centerline{
\includegraphics[width = \textwidth]{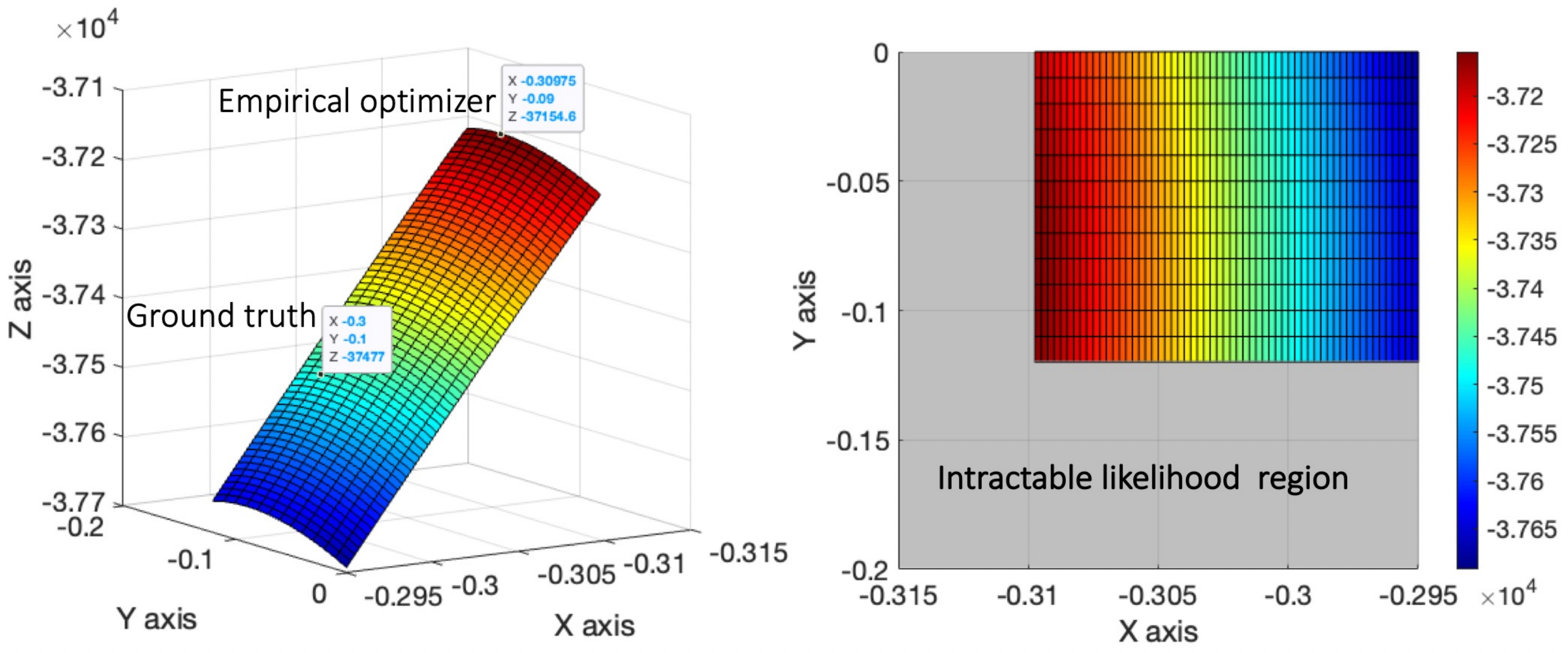}}
%\vspace{-0.1in}
\caption{Optimization landscape for a $d=3$ example; the X, Y and Z axes correspond to $\alpha_{13}$, $\alpha_{21}$ and the log likelihood, respectively. ``No Z value for pair (X, Y)'', which is the grey region in the right panel, means the log likelihood becomes intractable for the corresponding $(\alpha_{13}, \alpha_{21})$ pair. We can see the empirical optimizer lies on the border of the intractable likelihood region. Complete details of this illustrative example can be found in Appendix~\ref{appendix:d3example}.
}
\label{fig:opt_landscape}
%\vspace{-0.1in}
\end{figure}

\begin{figure*}[!htp]
%%%\vspace{-0.05in}
\centerline{
\includegraphics[width = \textwidth]{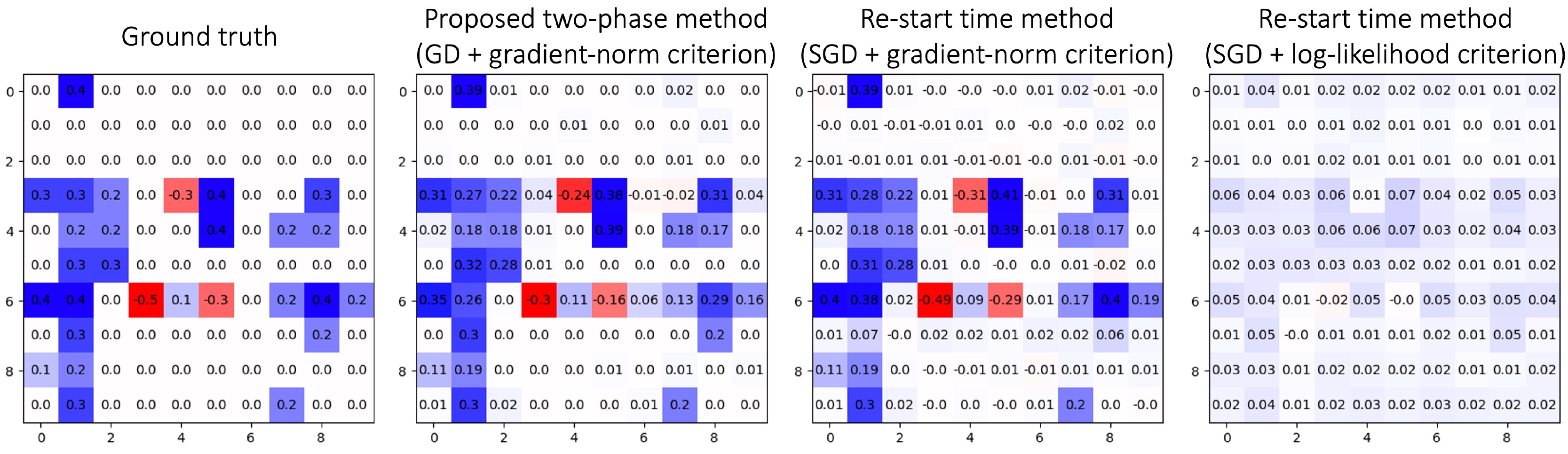}
}
%\vspace{-0.15in}
\caption{Comparison of the estimated adjacency matrices for a $d=10$ illustrative example; the corresponding method is specified on the top of each panel. We can observe that relaxing the feasibility constraint and using the gradient-norm as the stopping criterion can improve the estimation accuracy compared with the conventional likelihood criterion; further quantitative comparison can be found in Table~\ref{table:exp:baseline_single_seq}.}
\label{fig:comparison_illus}
%\vspace{-0.15in}
\end{figure*}

The difficulty of applying gradient descent (GD) comes from the optimization landscape --- the log likelihood can become {\it intractable}, i.e., the GD iterate could go outside the feasible region $\Theta$, especially when the it is close to the empirical optimizer as the empirical optimizer often lies on the edge of the feasible region (see Figure~\ref{fig:opt_landscape} for illustration), and the log likelihood will no longer be well-defined, rendering us unable to accurately track or maximize the likelihood to learn the problem parameters. This suggests that naively applying GD will result in a highly {\it unstable} procedure (as verified by Figure~\ref{fig:compare_GD_illus} in the appendix). 
Moreover, searching for the empirical optimizer within the feasible region based on the log likelihood criterion may not be the best option --- indeed, our empirical findings from Figure~\ref{fig:illus_criterion} in the appendix show that, even when the log likelihood becomes intractable, the estimation error continues to decrease when using the matrix Frobenius norm ($F$-norm) of the gradient with respect to (w.r.t.) adjacency matrix as the stopping criterion (referred to as the {\it gradient-norm criterion} below), suggesting that we could relax the {\it feasibility constraint} $(\mu, A) \in \Theta$ and use gradient-norm criterion instead of the log likelihood one.
To support this claim, we use SGD to solve for MLE within the feasible region \eqref{eq:feasible_region} and report the estimated adjacency matrix in the last panel in Figure~\ref{fig:comparison_illus}. In comparison, we relax the feasibility constraint and use the gradient-norm criterion. We report the resulting estimated $A$ in the third panel of Figure~\ref{fig:comparison_illus}, and we can see the estimation is more accurate when we use gradient-norm criterion compared with the conventional log likelihood criterion.

\begin{table}[!htp]
%%%\vspace{-0.2in}
\caption{Complexity analysis of different estimation methods; $d$ denotes the dimensionality and $N$ is the number of events. Since there is no adaptation of EM algorithm \citep{xu2016learning} to handle the instability issue as illustrated in Figure~\ref{fig:opt_landscape}, the gradient evaluation of EM is left blank.}\label{tab:complexity}
%\vspace{-0.15in}
\begin{center}
\begin{small}
\resizebox{0.75\textwidth}{!}{%
\begin{tabular}{lccc}
\toprule[1pt]\midrule[0.3pt]
 & EM & Re-start time & Proposed \\
\cmidrule(l){2-4}
Number of parameters & $O(N^2 + d^2)$ & $O(dN + d^2)$ & $O(d^2)$ \\
Gradient evaluation & $-$ & $O(dN^2)$ & $O(N^2 + dN)$ \\
\midrule[0.3pt]\bottomrule[1pt]
\end{tabular}
}
\end{small}
\end{center}
%\vspace{-0.2in}
\end{table}

\subsubsection{A maximum surrogate likelihood formulation}
Another practical issue comes from the re-start time \eqref{eq:re_start_time}, which needs to be re-calculated after each iteration, making it highly non-scalable; see Table~\ref{tab:complexity} for the complexity analysis. To alleviate this scalability issue caused by the re-start time calculation while harvesting the empirical good performance of \eqref{eq:re_start_lik}, we propose to maximize the following {\it surrogate log likelihood}:
\begin{align}
 \tilde \ell(\mu,A;\beta) =& \sum_{n = 1}^N \log  \tilde \lambda_{u_n}(t_n)- \sum_{i=1}^{d} \int_{0}^{T} \tilde \lambda_{i}(t) d t\label{eq:surrogate_lik}\\
 =&  \sum_{i=1}^d \sum_{j=1}^{N-1}  \frac{\alpha_{i,u_j}}{\beta} \left(e^{-\beta (t_{N}-t_j)} - 1\right) - T\sum_{i=1}^d \mu_i  + \sum_{n=1}^N  \log \bigg(\mu_{u_n} + \sum_{j < n} \alpha_{u_n,u_j} e^{-\beta (t_n-t_j)} \bigg), \nonumber
\end{align}
which serves as a computationally stable and efficient approximation to the true log likelihood \eqref{eq:re_start_lik}.
The above surrogate, which differs from true log likelihood in the integration region in the second term, is a computationally friendly and differentiable approximation to the true log likelihood, which can be understood as either (i) replacing the true conditional intensity with its differentiable surrogate \eqref{eq:surrogate_intensity} or (ii) ignoring the re-start time calculation and integrating the surrogate intensity on $[0,T]$. We will maximize this surrogate log likelihood to estimate the problem parameters, i.e.,
\begin{equation}\label{eq:MLE}
    \hat \mu, \hat A = {\rm argmin}_{(\mu,A) \in \Theta} - \tilde \ell(\mu,A;\beta).
\end{equation}

\subsubsection{A two-phase gradient descent algorithm}
\begin{figure*}[!htp]
%%%\vspace{-0.1in}
\centerline{
\includegraphics[width = .9\textwidth]{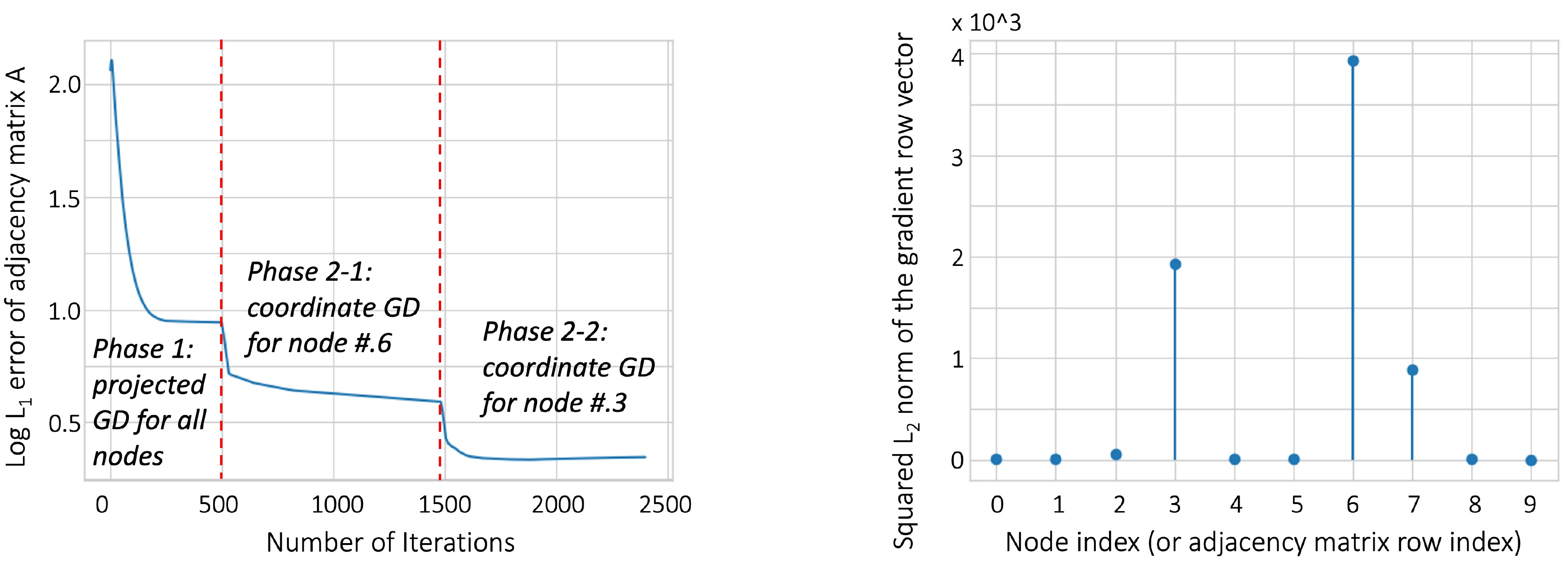}
}
%\vspace{-0.15in}
\caption{Illustration of the proposed two-phase method. After Phase 1, we select nodes \# 3 and \# 6 based on gradient-norm (see the right panel; the percentage threshold $p=0.85$ in Algorithm~\ref{algo:2_phase_2_index}) and proceed to phase 2, where we perform GD without projection for selected nodes (see the left panel for illustration and evidence of convergence).}
\label{fig:illus_2phase}
%\vspace{-0.15in}
\end{figure*}

Since the objective function \eqref{eq:MLE} is convex w.r.t. $(\mu, A)$ \citep{bacry2015hawkes}, projected gradient descent (PGD) is a tempting choice, which enjoys a strong convergence guarantee. However, despite the above simple closed-form expression, the projection back to $\Theta$ to maintain feasibility is computationally intense, making PGD again unscalable. 
Fortunately, the gradient field of this surrogate remains well-defined even outside the feasible region $\Theta$, making the vanilla GD possible.
However, vanilla GD (without projection) will suffer from divergence issues, as the iterate can easily go outside the feasible region $\Theta$ (see Figure~\ref{fig:compare_GD_illus} in the appendix for empirical evidence). 
Thus, we need to gradually decay the learning rate during the learning process.
Since the (surrogate) log likelihood is also intractable, it cannot be used to fulfill this purpose. To tackle those difficulties, we propose a two-phase GD-based method coupled with a learning rate decaying scheme based on the gradient-norm; this method is illustrated in Figure~\ref{fig:illus_2phase} and one can see its good performance in the second panel in Figure~\ref{fig:comparison_illus}.
Next, we will briefly introduce this algorithm.

\paragraph{Phase 1: Projected Gradient Descent.}
In the first phase, we constrain all parameters to be non-negative and perform projected GD with fixed step length. We denote $\hat \mu_t$ and $\hat A_t$ to be the iterates at $t$-th step, and the update rule is as follows:
\begin{align*}
\hat \mu_t  \leftarrow \hat \mu_{t-1} + \gamma \nabla_{\mu} \tilde \ell/\|\nabla_{\mu} \tilde \ell\|_2, \quad \hat A_t  \leftarrow \hat A_{t-1} + \gamma \nabla_{A} \tilde \ell/\|\nabla_{A} \tilde \ell \|_F,
\end{align*}
where $\gamma$ is the learning rate, $\norm{\cdot}_2$ and $\norm{\cdot}_F$ represent vector $L_2$ norm and matrix Frobenius norm, respectively, and the gradient fields are defined as:
$$\nabla_{\mu} \tilde \ell = \nabla_{\mu} \tilde \ell(\hat \mu_{t-1},\hat A_{t-1};\beta), \quad   \nabla_{A} \tilde \ell = \nabla_{A} \tilde \ell(\hat \mu_{t-1},\hat A_{t-1};\beta).$$ 
The parameter $\beta$ is assumed to be known; in practice, we will perform a grid search to select the best $\beta$. To make sure we do not get negative intensity, we perform the following projection:
\begin{align*}
        \hat \mu_t \leftarrow {\rm argmin}_{\mu \in \mathbb{R}_+} \norm{\hat \mu_t - \mu}_2, \quad\hat A_t \leftarrow {\rm argmin}_{A \in \mathbb{R}_+^{d\times d}} \norm{\hat A_t - A}_F,
\end{align*}
where $\mathbb{R}_+ = [0,\infty)$.
This projection can be easily achieved by setting all negative entries to zeros; complete details of the PGD can be found in Algorithm~\ref{algo:2_phase_1} in Appendix~\ref{appendix:add_alg}.

This warm-up phase guides us to a neighborhood around the global optimizer while ensuring the stability/convergence of the algorithm. 
%(see a illustrative example of vanilla GD performance in Figure~\ref{fig:compare_GD_illus}) 
Moreover, it reduces the computation cost by finding a small batch of coordinates for further optimization in phase 2; see the illustration in Figure~\ref{fig:illus_2phase} and the description of the phase 2 algorithm below.

%%%%\vspace{0.05in}

\paragraph{Phase 2: Batch Coordinate Gradient Descent.} In the second phase, we consider those variables/nodes whose corresponding rows (in $A$) could have negative values. We identify those nodes by the $L_2$ norm of the gradient (w.r.t. $A$) row vector --- large gradient-norm indicates that the convergence of the corresponding row is not achieved yet after phase 1; see the right panel in Figure~\ref{fig:illus_2phase} for a graphical illustration and complete details on how to identify those rows in Algorithm~\ref{algo:2_phase_2_index} in Appendix~\ref{appendix:add_alg}. 
Next, we need to keep performing GD without the constraint/projection for those selected rows in $A$ to estimate those negative entries (and PGD for the corresponding background intensities).
Despite the intractable log likelihood in this phase, we develop a learning rate decaying scheme based on the gradient $F$-norm to guarantee convergence empirically. Complete details of the PGD algorithm can be found in Algorithm~\ref{algo:2_phase_2_main} in Appendix~\ref{appendix:add_alg}.

Recently, \citet{juditsky2019signal,juditsky2020convex} showed that a projected GD along some (strong) monotone vector field can be interpreted as a solution to a stochastic variation inequality (VI) and enjoys both signal estimation guarantee and convergence guarantee. 
However, since we do not constraint the iterate within $\Theta$ in phase 2, the vector fields $\nabla_{\mu} \tilde \ell$ and $\nabla_{A} \tilde \ell$ are no longer monotone. Hence, we could only use numerical evidence to show the effectiveness of our method.
Nevertheless, this vector field view under the VI framework might give us a chance to theoretically explain our heuristic's empirical success.

\section{Experiments}

%%%\vspace{-0.1in}
\subsection{Numerical simulation}\label{sec:exp}
%%%\vspace{-0.1in}

In this subsection, we will show the good performance of our proposed two-phase method. We report (i) $L_1$ norm of $\beta$ estimation error (\texttt{$\beta$ err.}), 
(ii) $L_1$ norm of $\mu$ estimation error (\texttt{$\mu$ err.}), (iii) $L_1$ norm of $A$ estimation error (\texttt{$A$ err.}), 
(iv) Hamming Distance (\texttt{$A$ HD}) and (v) Structural Hamming Distance (\texttt{$A$ SHD}) between ground truth and estimated adjacency matrix $A$ as our evaluation metrics. All experiments in this subsection are carried out for randomly generated problem parameters and repeated 100 times; here we report the mean and standard deviation of those metrics.
One can see Appendix~\ref{appendix:exp_detail} for further details.

\subsubsection{Experiment 1} We begin with a simple setting where we know the ground truth $\beta$. We want to numerically verify the consistency with respect to the time horizon $T$ and the total number of sequences.
To be precise, we generate (1) one single sequence on time horizon $T \in \{500,2000,5000,$ $10000,20000\}$ and (2) multiple sequences (total sequence number chosen from $\{1,10,20,50,100\}$) on time horizon $T = 500$ and learn the parameter via our proposed two-phase method. 
We report the results for $d=5, 10$ cases in Table~\ref{table:exp:T_consistency}, from which we can see that, with longer sequences (or more sequences), all those errors decrease monotonically.
To further validate our findings, we also perform the experiment for $d=20$ case; the results can be found in Table~\ref{table:exp:consistency_d20} in Appendix~\ref{appendix:add_exp}, from which we can still see the decaying error pattern as observed in the above $d=5, 10$ cases. Therefore, we numerically verify the consistency of our proposed method.

\begin{table*}[!htp]
{\caption{Performance of proposed method when $\beta$ is assumed to be known. We observe that all error metrics are decreasing with either an increasing number of sequences or time horizon, which numerically verifies the consistency of our method.}\label{table:exp:T_consistency}}%
\vspace{0.2in}
\resizebox{\textwidth}{!}{%
\begin{tabular}{lcccccccccccccr}
 \multicolumn{11}{c}{\large{Varying sequence time horizon $T$ (sequence number fixed to be 1).}} \\ 
\toprule[1pt]\midrule[0.3pt]
& \multicolumn{5}{c}{\large{$d=5$}} & \multicolumn{5}{c}{\large{$d=10$}}\\
T & 500 & 2000 & 5000 & 10000 & 20000 & 500 & 2000 & 5000 & 10000 & 20000 \\% Column names row
\cmidrule(l){2-6}
\cmidrule(l){7-11}
 % In-table horizontal line
\texttt{$\mu$ err.$^\star$} & 7.41  (3.42) & 5.34  (2.93) & 4.25  (2.76) & 3.81  (2.8) & 3.69  (2.72) & 17.96  (5.98) & 11.26  (4.47) & 8.65  (4.01) & 8.78  (3.8) & 7.58  (3.75)\\
\texttt{$A$ err.} & 8.94 (5.96) & 2.26 (2.81) & 1.01 (0.82) & 0.75 (0.43) & 0.57 (0.23) & 20.95 (12.55) & 4.93 (2.55) & 2.6 (1.04) & 1.87 (0.61) & 1.52 (0.48)  \\
\texttt{$A$ HD} & 0.24 (0.13) & 0.08 (0.091) & 0.04 (0.076) & 0.04 (0.059) & 0.0 (0.056) & 0.245 (0.075) & 0.07 (0.063) & 0.03 (0.035) & 0.015 (0.025) & 0.01 (0.018)\\
\texttt{$A$ SHD} & 6.0 (3.43) & 2.0 (2.28) & 1.0 (1.9) & 1.0 (1.48) & 0.0 (1.41)& 24.5 (7.72) & 7.0 (6.34) & 3.0 (3.57) & 1.5 (2.55) & 1.0 (1.86) \\
\midrule[0.3pt]
\bottomrule[1pt]
%\multicolumn{5}{l}{$\star$ we omit $\times 10^{-2}$ in the value due to space consideration.} \\
\end{tabular}
}

\rule{0pt}{.15in} \\

\resizebox{\textwidth}{!}{%
\begin{tabular}{lcccccccccccccr}
 \multicolumn{11}{c}{\large{Varying sequence number (time horizon $T$ fixed to be $500$).}} \\
\toprule[1pt]\midrule[0.3pt]
& \multicolumn{5}{c}{\large{$d=5$}} & \multicolumn{5}{c}{\large{$d=10$}}\\
Seq. Num. & 1 & 10 & 20 & 50 & 100 & 1 & 10 & 20 & 50 & 100 \\% Column names row
\cmidrule(l){2-6}
\cmidrule(l){7-11}
 % In-table horizontal line
\texttt{$\mu$ err.$^\star$} & 6.25  (3.29) & 3.91  (2.81) & 3.86  (2.72) & 3.41  (2.66) & 2.91  (2.50) & 17.96  (5.98) & 8.61  (4.01) & 8.67  (3.78) & 7.54  (3.74) & 6.9  (3.49)\\
\texttt{$A$ err.} & 9.42 (5.50) & 1.19 (1.22) & 0.86 (1.00) & 0.6 (0.91) & 0.54 (0.91) & 20.96 (12.56) & 2.62 (1.04) & 1.84 (0.61) & 1.4 (0.46) & 1.51 (0.47) \\
\texttt{$A$ HD} & 0.26 (0.120) & 0.06 (0.075) & 0.04 (0.061) & 0.04 (0.045) & 0.0 (0.050) & 0.245 (0.075) & 0.03 (0.036) & 0.015 (0.026) & 0.01 (0.018) & 0.01 (0.017)\\
\texttt{$A$ SHD} & 7.0 (3.14) & 1.5 (1.89) & 1.0 (1.53) & 1.0 (1.14) & 0.0 (1.27)& 24.5 (7.72) & 3.0 (3.66) & 1.5 (2.71) & 1.0 (1.85) & 1.0 (1.76)  \\
\midrule[0.3pt]
\bottomrule[1pt]
\multicolumn{11}{l}{$\star$ the value times $10^{-2}$ is the actual $\mu$ estimation error; we omit $\times 10^{-2}$ in the value due to space consideration.}
\\
\end{tabular}
}
%\vspace{-0.1in}
\end{table*}

\subsubsection{Experiment 2} Next, we consider a more general scenario where we do not know the true $\beta$ (ground truth is $0.8$) --- we treat it as a hyperparameter and perform a grid search over $\beta \in \{0.4, 0.5, 0.6, 0.7, $ $0.8, 0.9,$ $1, 1.1, 1.2\}$. We propose to use the {\it end-of-phase 1 likelihood} as the goodness-of-fit (GoF) criterion to select hyperparameter $\beta$. In comparison, we also consider the end-of-phase 1 gradient-norm as the GoF criterion, but it does not perform as well as the end-of-phase 1 likelihood criterion; one can see the corresponding results in Table~\ref{table:exp:beta_grid_saerch_grad} in Appendix~\ref{appendix:add_exp}.

For each grid value, we randomly generate synthetic data ($50$ sequences with $T=500$) and fit our model. We repeat this procedure independently 100 times, and at each trial, we select $\beta$ with the largest end-of-phase 1 likelihood.
We report the results in Table~\ref{table:exp:beta_grid_saerch_lik}, where we can observe that the grid search approach coupled with end-of-phase 1 likelihood GoF criterion achieves almost the same performance with the best achievable performance (oftentimes it is better than true $\beta$'s performance). This shows the effectiveness of our approach in practice.
% ; second, from the $A$ error behavior in Table~\ref{table:exp:beta_grid_saerch_lik}, we can also verify the convexity of the objective function w.r.t. $\beta$.

\begin{table*}[!htbp]
%%%\vspace{-0.1in}
%\centering
{\caption{Performance of proposed method when $\beta$ is unknown. The last row corresponds to selected $\beta$ based on end-of-phase 1 log likelihood, where we can observe its performance (italic) is almost the same with the best achievable performance (bold).}\label{table:exp:beta_grid_saerch_lik}}
\vspace{0.2in}
\resizebox{1\textwidth}{!}{%
\begin{tabular}{lcccccccccccccr}
\toprule[1pt]\midrule[0.3pt]
& \multicolumn{4}{c}{$d=5$} & \multicolumn{4}{c}{$d=10$} & \multicolumn{4}{c}{$d=20$} \\
$\beta$ & \texttt{$\mu $ err.$^\star$} & \texttt{$A$ err.} & \texttt{$A$ HD} & \texttt{$A$ SHD} & \texttt{$\mu $ err.$^\star$} & \texttt{$A$ err.} & \texttt{$A$ HD} & \texttt{$A$ SHD} & \texttt{$\mu $ err.$^\star$} & \texttt{$A$ err.} & \texttt{$A$ HD} & \texttt{$A$ SHD} \\
\cmidrule(l){2-5}
\cmidrule(l){6-9}
\cmidrule(l){10-13}
        0.4 & 5.01   (3.16) & 1.51 (0.85) & 0.06 (0.092) & 1.5 (2.3) & 8.53   (3.81) & 4.39 (0.63) & 0.03 (0.052) & 3.0 (5.27) & 14.97   (5.13) & 13.46 (2.14) & 0.047 (0.044) & 19.0 (17.82) \\ 
        0.5 & 5.78   (3.3) & 1.26 (0.86) & 0.04 (0.08) & 1.0 (2.01) & 10.57   (4.39) & 3.49 (0.6) & 0.02 (0.033) & 2.0 (3.39) & 20.51   (6.16) & 10.59 (2.27) & \textbf{0.043 (0.047)} & \textbf{17.5 (18.88)} \\ 
        0.6 & 5.39   (3.24) & 1.05 (0.88) & 0.02 (0.071) & 0.5 (1.79) & 10.04   (4.36) & 2.58 (0.64) & 0.02 (0.023) & 2.0 (2.38) &  21.18   (6.41) & 8.55 (2.35) & 0.045 (0.046) & 18.0 (18.67)\\
        0.7 & 5.2   (3.12) & 0.86 (0.88) & 0.0 (0.065) & 0.0 (1.64) & 8.94   (4.05) & 1.86 (0.61) & 0.01 (0.024) & 1.0 (2.47) & 19.35   (6.09) & 6.54 (2.43) & 0.048 (0.048) & 19.5 (19.24)\\ 
        0.8 & 4.67   (3.02) & 0.74 (0.89) & 0.0 (0.039) & 0.0 (0.99) & 7.54   (3.74) & \textbf{1.4 (0.46)} & \textbf{0.01 (0.018)} & \textbf{1.0 (1.85)} & 17.11   (5.65) & \textbf{4.98 (2.5)} & 0.06 (0.053) & 24.0 (21.52)\\ 
        0.9 & 4.51   (2.94) & \textbf{0.66 (0.9)} & 0.0 (0.036) & 0.0 (0.91) & \textbf{6.79   (3.53)} & 1.52 (0.41) & 0.01 (0.02) & 1.0 (2.08) & \textbf{16.34   (5.21)} & 5.11 (2.13) & 0.07 (0.055) & 28.0 (22.32)\\ 
        1 & \textbf{4.46   (2.87)} & 0.79 (0.92) & 0.0 (0.034) & 0.0 (0.86) & 7.16   (3.47) & 1.84 (0.39) & 0.01 (0.022) & 1.0 (2.28) & 18.0   (5.29) & 5.93 (1.82) & 0.088 (0.055) & 35.5 (22.29)\\ 
        1.1 & 4.56   (2.79) & 1.03 (0.91) & 0.0 (0.035) & 0.0 (0.87) & 7.73   (3.47) & 2.3 (0.34) & 0.02 (0.026) & 2.0 (2.65) & 19.95   (5.55) & 6.83 (1.64) & 0.103 (0.056) & 41.5 (22.52)\\ 
        1.2 & 4.75   (2.74) & 1.2 (1.22) & \textbf{0.0 (0.032)} & \textbf{0.0 (0.85)} & 8.4   (3.46) & 2.72 (0.38) & 0.03 (0.033) & 3.0 (3.36)  & 22.84   (6.08) & 7.8 (1.53) & 0.121 (0.052) & 48.5 (21.1)\\ 
        $-$ & {\it 4.57   (2.96)} & {\it0.74 (0.89)} & {\it 0.0 (0.036)} & {\it 0.0 (0.9)} & {\it 7.04   (3.55)} & {\it 1.62 (0.41)} & {\it 0.01 (0.021)} & {\it 1.0 (2.17)} & {\it 16.7   (5.34)} & {\it 5.06 (2.19)} & {\it 0.07 (0.055)} & {\it 28.0 (22.02)}\\ 
\midrule[0.3pt]
\bottomrule[1pt]
\multicolumn{12}{l}{$\star$ the value times $10^{-2}$ is the actual $\mu$ estimation error; we omit $\times 10^{-2}$ in the value due to space consideration.}
\\
\end{tabular}
}
%\vspace{-0.15in}
\end{table*}

\subsubsection{Experiment 3}\label{subsec:exp3_comparison_benchmark}  
Lastly, we compare our proposed method with two benchmark methods for $d=20$ setting. 
Here, we consider vanilla gradient descent (vanilla GD) and the re-start time method \citep{bonnet2021maximum,bonnet2022inference} coupled with SGD (gradient-norm as the stopping criterion); further details of the benchmarks can be found in Appendix~\ref{appendix:benchmark}. We report the results in Table~\ref{table:exp:baseline_single_seq}. As we can see, for both cases, our proposed method outperforms benchmarks in terms of most evaluation metrics; in particular, our method does the best in terms of adjacency matrix recovery. Although GD with proper stopping criterion achieves slightly better $L_1$ estimation error in case 1, its pattern recovery of $A$ is much worse than our method (i.e., larger HD and SHD), making it unable to return a reliable GC graph in practice. 

\begin{table}[!htp]
%\vspace{-0.1in}
\caption{Comparison with benchmarks. The best results are highlighted. We can observe our method performs the best in terms of the adjacency matrix recovery.}\label{table:exp:baseline_single_seq}
%\vspace{-0.15in}
\begin{center}
\begin{small}
\resizebox{.75\textwidth}{!}{%
\begin{tabular}{lcccc}
\multicolumn{5}{c}{{Case 1: single sequence with time horizon $T=10000$.}} \\ 
\toprule[1pt]\midrule[0.3pt]
Method & Two-phase method & Vanilla GD & Early stopped GD & Re-start  \\% Column names row
\cmidrule(l){2-5}
 % In-table horizontal line
\texttt{$\beta$ err.} & .312 (.112) & .393 (.035) & \textbf{.264} (.137) & .837 (.246) \\
\texttt{$\mu$ err.} & \textbf{.0386} (.0252) & .0413 (.0317) & .0398 (.0281) & .239 (.102) \\
\texttt{$A$ err.} & 1.726 (0.785) & 23.58 (7.93) & \textbf{1.494} (0.731) & 8.828 (1.213)  \\
\texttt{$A$ HD} & \textbf{.0304} (.0416) & .1336 (0.118) & .0936 (.0926) & .3576 (.0459)\\
\texttt{$A$ SHD} & \textbf{.76} (1.04) & 3.37 (2.96) & 2.34 (2.32) & 8.98 (1.19) \\
\midrule[0.3pt]\bottomrule[1pt]
\end{tabular}
}
\end{small}
\end{center}

%%\vspace{-0.1in}

\begin{center}
\begin{small}
\resizebox{0.75\textwidth}{!}{%
\begin{tabular}{lcccc}
\multicolumn{5}{c}{{Case 2: multiple (100) sequences with time horizon $T = 500$.}} \\ 
\toprule[1pt]\midrule[0.3pt]
Method & Two-phase method & Vanilla GD & Early stopped GD & Re-start + SGD  \\% Column names row
\cmidrule(l){2-5}
 % In-table horizontal line
\texttt{$\beta$ err.} & {.264} (.151) & .367 (.074) & \textbf{.254} (.136) & .295 (.128) \\
\texttt{$\mu$ err.} & .0489 (.0269) & .0367 (.0228) & .0435 (.0272) & \textbf{.0198} (.0136) \\
\texttt{$A$ err.} & \textbf{0.983} (0.301) & 12.13 (2.419) & 1.759 (0.480) & 1.067 (0.401)   \\
\texttt{$A$ HD} & \textbf{.0236} (0.0376) & .0748 (0.0680) & .0808 (0.0642) & .044 (0.0639)\\
\texttt{$A$ SHD} & \textbf{.59} (.94) & 1.88 (1.72) & 2.02 (1.61) & 1.1 (1.60) \\
\midrule[0.3pt]\bottomrule[1pt]
\end{tabular}
}
\end{small}
\end{center}
%\vspace{-0.1in}
\end{table}

Additionally, as shown in Table~\ref{tab:complexity}, our proposed approach is scalable, which is another major advantage compared with the re-start time approach. Here, we demonstrate this benefit by performing a run time analysis. Due to space consideration, the results are deferred to Table~\ref{table:exp:runtime} in Appendix~\ref{appendix:runtime}.

%%%\vspace{-0.2in}
\subsection{Real data example}\label{sec:real_data_exp}

We created a retrospective cohort of patients utilizing in-hospital data derived from Grady hospital system in Atalanta, GA, an academic level 1 trauma center, 
spanning 2018-2019. This data was collected and analyzed in accordance with Emory Institutional Review Board (IRB) approved protocol \#STUDY00000302.
Patients were included in the Sepsis-3 cohort if they met Sepsis-3 criteria while in the hospital and were admitted for $\geq$ 24 hours. Patients were included in the Non-Septic cohort if they had a Sequential Organ Failure Assessment (SOFA) score $\geq$2. 
A total of 37 patient features comprised of laboratory results (Labs) and observations (vital signs) were examined for this work. Treatments were limited to two classes of medication: antimicrobial therapy (e.g., antibiotics) and vasopressor therapy. We report the median and interquartile range (IQR) in Table~\ref{table:demo} and defer the cohort construction details to Appendix~\ref{appendix:lab_vital_SAD}. 
\begin{table}[htp]
%\vspace{-0.1in}
\caption{Summary statistics of patients' demographics.}\label{table:demo}
%\vspace{-0.15in}
\begin{center}
\begin{small}
\resizebox{0.85\textwidth}{!}{%
\begin{tabular}{lcccc}
\toprule[1pt]\midrule[0.3pt]
& \multicolumn{2}{c}{{Sepsis-3 patients}} & \multicolumn{2}{c}{{Non-sepatic patients}} \\ 
year & 2018 ($n=409$) $^\star$ & 2019 ($n = 454$) & 2018 ($n = 960$) & 2019 ($n = 1169$)  \\% Column names row
\cmidrule(l){2-5}
 % In-table horizontal line
{Age (median and IQR)} & 58 (38 - 68) & 59 (46 - 68) & 56 (38 - 67) & 55 (37 - 66) \\
{Female (percentage)} & 30.1 $\%$ & 36.6 $\%$ & 37.1 $\%$ & 35.8 $\%$ \\
{SOFA score (mean)} & 3.32& 3.14 & 2.18 & 2.28  \\
% \texttt{Total vent. days} & 4 (2 - 8) & 4 (3 - 8) & 3 (2 - 6) & 3 (2 - 5) \\
% \texttt{Total ICU days} & 5 (3 - 8) & 5 (3 - 8) & 3 (2 - 6) & 3 (2 - 6)\\
% \texttt{Total hosp. days} & 11.5 (6 - 20) & 11 (6 - 18) & 10 (5 - 17) & 8 (4 - 16) \\
{Traj. len. (median and IQR)} & 25 (25 - 25) & 25 (25 - 25) & 17 (13 - 22) & 17 (13 - 22) \\
\midrule[0.3pt]\bottomrule[1pt]
\multicolumn{5}{l}{$\star$ $n$ represents the total number of patients in the corresponding cohort.}
\end{tabular}
}
\end{small}
\end{center}
%\vspace{-0.15in}
\end{table}

\subsubsection{Sepsis-Associated Derangements}
Integrating high dimensional information (via, e.g., clustering) is essential in causal discovery and explainable machine learning \citep{sanchez2022causal}; examples include \citet{uleman2021mapping,braman2021deep,wei2021inferringb}. While the Sepsis-3 definition provides the explicit features necessary for identifying the presence of sepsis, there is no consensus as to which features are best for prognosticating the disease. To reduce the complexity of our computations, expert opinion was utilized to identify common and clinically relevant Sepsis-Associated Derangements (SADs) that could be detected using structured EMR data. A total of $18$ SADs and $2$ relevant treatments shown in Table~\ref{table:SADname} were identified using 37 patient features and treatments gathered from the medical record. A SAD was considered present if the patient's features were outside of normal limits. Details on how SADs were constructed based on vital signs and Labs can be found in Table~\ref{table:SADcutoff} in Appendix~\ref{appendix:lab_vital_SAD}.

\begin{table}[!htp]
%\vspace{-0.1in}
\caption{Measurements in sepsis-associated events construction.}\label{table:SADname}
%\vspace{-0.2in}
\begin{center}
\begin{small}
\resizebox{.9\textwidth}{!}{%
\begin{tabular}{lll}
\multicolumn{3}{c}{\Large{Sepsis-Associated Derangement}}\\
\toprule[1pt]\midrule[0.3pt]
Full name &  Abbreviation & Measurement name \\ % Column names row
\midrule[0.3pt]
\textbf{Renal Dysfunction} & RenDys &creatinine, blood\_urea\_nitrogen\_(bun)\\
\textbf{Electrolyte Imbalance} & LyteImbal &calcium, chloride, magnesium, potassium, phosphorus\\
\textbf{Oxygen Transport Deficiency} & O2TxpDef &hemoglobin\\
\textbf{Coagulopathy} & Coag &partial\_prothrombin\_time\_(ptt), fibrinogen, platelets, \\
& &  d\_dimer, thrombin\_time, prothrombin\_time\_(pt), inr\\
\textbf{Malnutrition} & MalNut &transferrin, prealbumin, albumin\\
\textbf{Cholestatsis} & Chole &bilirubin\_direct, bilirubin\_total\\
\textbf{Hepatocellular Injury} & HepatoDys &aspartate\_aminotransferase\_(ast), ammonia,\\
& &  alanine\_aminotransferase\_(alt)\\
\textbf{Acidosis} & Acidosis &base\_excess, ph\\
\textbf{Leukocyte Dysfunction} & LeukDys &white\_blood\_cell\_count\\
\textbf{Hypercarbia} & HypCarb & partial\_pressure\_of\_carbon\_dioxide\_(paco2),\\
& & end\_tidal\_co2\\
\textbf{Hyperglycemia} & HypGly &glucose\\
\textbf{Mycardial Ischemia} & MyoIsch &troponin\\
\textbf{Tissue Ischemia} & TissueIsch &base\_excess, lactic\_acid\\
\textbf{Diminished Cardiac Output} & DCO &best\_map\\
\textbf{CNS Dysfunction} & CNSDys &gcs\_total\_score\\
\textbf{Oxygen Diffusion Dysfunction} & O2DiffDys &spo2, fio2\\
\textbf{Thermoregulation Dysfunction} & ThermoDys &temperature\\
\textbf{Tachycardia} & Tachy &pulse\\
\midrule[0.3pt]\bottomrule[1pt]

\rule{0pt}{.51ex} \\

 % In-table horizontal line
\multicolumn{3}{c}{\Large{Other Sepsis-Associated Events}}\\
\toprule[1pt]\midrule[0.3pt]
Full name &  Abbreviation & Measurement name \\ % Column names row
\midrule[0.3pt]
\textbf{Vasopressor Support} & VasoSprt &norepinephrine\_dose\_weight, epinephrine\_dose\_weight, \\
& & dobutamine\_dose\_weight, dopamine\_dose\_weight, \\
& &  phenylephrine\_dose\_weight, vasopressin\_dose\_weight\\
\textbf{Antibiotic Therapy} & ABX & $-$ \\
\textbf{Sepsis} & SEP3 & $-$\\
\midrule[0.3pt]\bottomrule[1pt]
\end{tabular}
}
\end{small}
\end{center}
%\vspace{-0.2in}
\end{table}

\begin{figure}[!htp]
%%%\vspace{-0.05in}
\centerline{
\includegraphics[width = .85\textwidth]{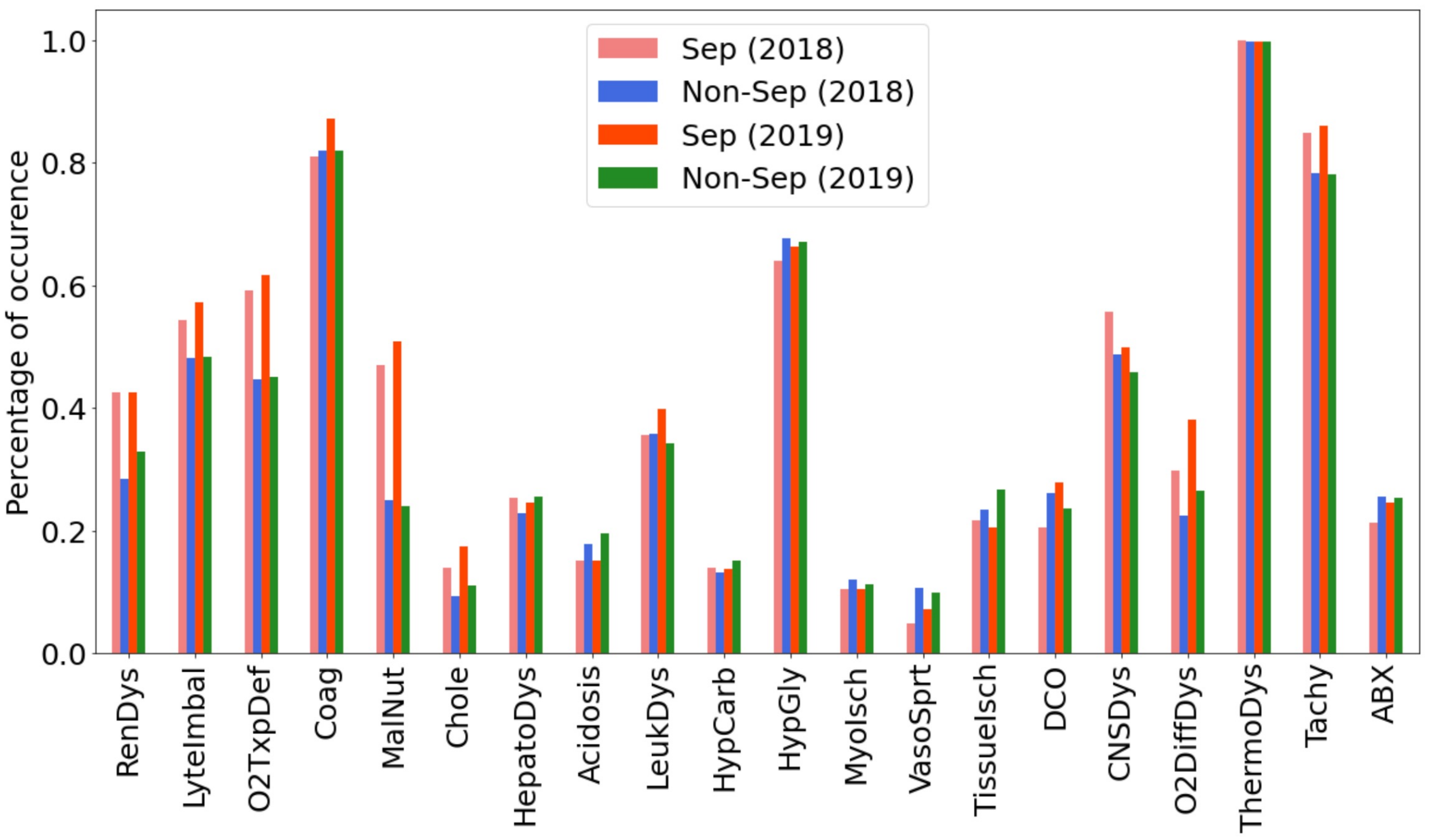}}
%\vspace{-0.13in}
\caption{Percentage of SAD's occurrence. We can observe that most SADs' occurrences are more frequent in Sepsis-3 cohort, which can justify our approach to construct SADs. 
}
\label{fig:SAD_percent}
%\vspace{-0.22in}
\end{figure}

Sepsis often shares symptoms with other disease processes making discrimination challenging. To evaluate the appropriateness of the constructed SADs, the percentage of SAD occurrence (within the selected time window) was calculated for both Sepsis and Non-Septic patients and can be seen for both years in Figure~\ref{fig:SAD_percent}. It is expected that SADs would be present in both cohorts; however, the Sepsis-3 cohort demonstrated patterns showing a closer relationship with the SADs than the Non-Septic cohort.

\subsubsection{Recovering GC graphs}
\begin{figure*}[!bhtp]
%%%\vspace{-0.1in}
\centerline{
\includegraphics[width = .49\textwidth]{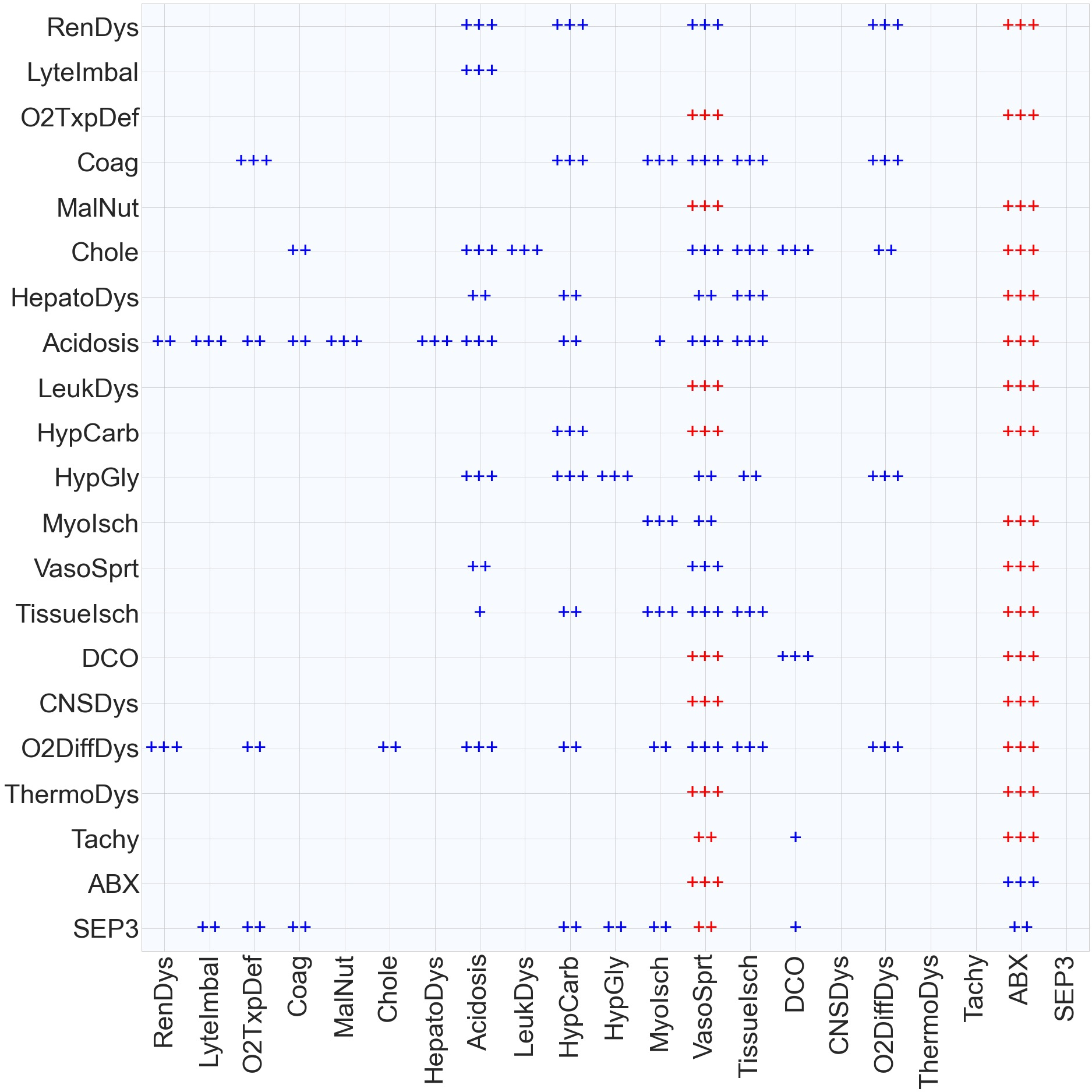}

\hspace{0.1in}

\includegraphics[width = .49\textwidth]{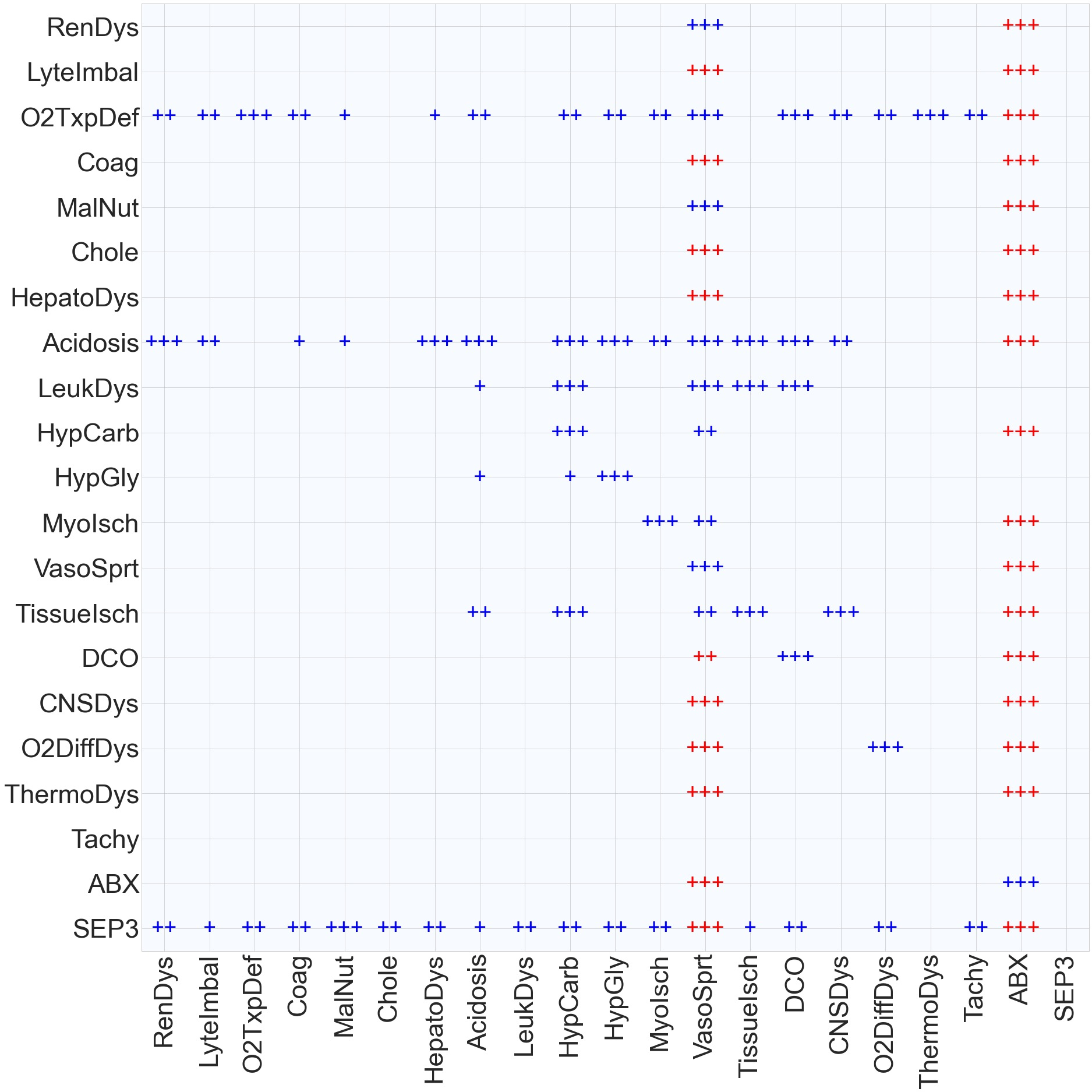}
}
%\vspace{-0.13in}
\caption{Adjacency matrices of the Granger Causal graphs for Sepsis-3 (left) and full (right) patient cohorts in Figure~\ref{fig:graph}. ``+'', ``++'' and ``+++'' correspond to the (absolute) value in $(0,.0005)$, $[.0005,.001)$ and $[.001,\infty)$, respectively, where the original values in the adjacency matrices are reported in Figure~\ref{fig:mat_num} in the Appendix. Nodes (i.e., SADs or SAEs) named along the X-axis can have either an inhibitory (red) or promoting effect (blue) on the nodes named on the Y-axis.} 
\label{fig:mat_plus}
%\vspace{-0.13in}
\end{figure*}

To study the temporal interactions between SADs (and other SAEs), we fit two GC graphs --- one graph is on the Sepsis-3 cohort and the other is on the full patient cohort (i.e., the Sepsis-3 and Non-Septic cohorts combined). We report the results in Figure~\ref{fig:mat_plus} and defer the training details to Appendix~\ref{appendix:realdata_GC}. Both graphs demonstrate examples of clinically reasonable interactions between individual SADs (i.e., Oxygen Diffusion Dysfunction promotes Renal Dysfunction in the Septic cohort) and between SADs and Sepsis (i.e., Diminished Cardiac Output promotes Sepsis in both graphs). Interestingly the graph examining only the Sepsis-3 cohort identified more interactions between SADs than the one for the full patient cohort whereas the graph for the full patient cohort presented a higher number of strong relationships between SADs and sepsis suggesting that a time-dependent, causal relationship exists between individual SADs and sepsis. A key finding across both graphs was the inhibitory effect of antibiotics on most SADs, which is consistent with the known ability of antibiotics to reduce in-hospital mortality in sepsis patients \citep{seymourTimeTreatmentMortality2017} presumably through preventing organ dysfunction like those identified via SADs. 

While most of the relationships identified in these graphs are expected or feasible, vasopressors appear to unexpectedly inhibit both sepsis and the administration of antibiotics. In the year 2018, among 409 (960) selected septic (non-septic) patients, there were 15 (96) receive vasopressor support and 84 (231) received antibiotics during the window, and only 3 (38) received both vasopressors and antibiotics. This low number of vasopressor patients in the Sepsis-3 cohort is not unexpected as the time window for analysis is 24 hours prior to meeting the Sepsis-3 definition when most patients are not severely ill (see Appendix~\ref{appendix:lab_vital_SAD} for more details). Additionally, antibiotics are dosed at scheduled intervals (e.g., once every six hours) whereas vasopressors are administered in a continuous fashion. These two attributes of the data set create a number of instances where vasopressors are administered without a formal antibiotic administration event in the following hour (though the patient may be on antibiotics). Additionally, each patient in the Sepsis-3 cohort is right censored after sepsis which means there is only one hour where the sepsis label is positive. Taken together these attributes of the data set likely explain why this unexpected relationship is seen.

\subsubsection{Identifying GC chains}
The estimated GC graphs help reduce the problem of enumerating combinatorially many possible chains to find the chains that only exist in the Sepsis-3 graph.
However, even for a 2-by-2 sub-adjacency matrix, there could be multiple potential chain interpretations. We validate whether or not the chain structure reflects a unique pattern in the Sepsis-3 cohort by performing Fisher's exact test and reporting the $p$-value. Here, we only focus on the ``++'' and ``+++'' exciting effects when forming all possible chains. This method allows chains to be ranked in order of significance, affording those with domain expertise an efficient mechanism to inspect results. We report the top GC chains which are unique in the Sepsis-3 cohort for years 2018 (in-sample test) and 2019 (out-of-sample test) in Table~\ref{table:chain}. More details on those chains (including how to perform the test) and more identified chains can be found in Tables~\ref{table:chain1}, \ref{table:chain2} and \ref{table:chain3} 
in Appendix~\ref{appendix:realdata_chain}.

\begin{table}[!htp]
%\vspace{-0.05in}
\caption{Granger Causal chains which are significantly unique in Sepsis-3 cohort in both years 2018 and 2019.}\label{table:chain}
\vspace{0.1in}
\begin{center}
\begin{small}
\resizebox{0.6\textwidth}{!}{%
\begin{tabular}{crcccc}
\toprule[1pt]\midrule[0.3pt]
Chain: & TissueIsch  & $\rightarrow$ &  O2DiffDys \\ 
$p$-value: & \multicolumn{2}{c}{{ 0.004 (2018)}} & \multicolumn{2}{c}{{ 0.092 (2019)}} \\ 
\cmidrule(l){1-6}
Chain: & O2DiffDys  & $\rightarrow$ &  RenDys & $\rightarrow$ &  O2DiffDys \\ 
$p$-value: & \multicolumn{2}{c}{{ 0.107 (2018)}} & \multicolumn{2}{c}{{ 0.004 (2019)}} \\ 
\cmidrule(l){1-6}
Chain: & VasoSprt  & $\rightarrow$ &  TissueIsch & $\rightarrow$ &  HepatoDys \\ 
$p$-value: & \multicolumn{2}{c}{{ 0.052 (2018)}} & \multicolumn{2}{c}{{ 0.088 (2019)}} \\ 
\cmidrule(l){1-6}
Chain: & LyteImbal  & $\rightarrow$ &  Acidosis & $\rightarrow$ &  O2DiffDys \\ 
$p$-value: & \multicolumn{2}{c}{{ 0.009 (2018)}} & \multicolumn{2}{c}{{ 0.088 (2019)}} \\ 
\cmidrule(l){1-6}
Chain: & Acidosis  & $\rightarrow$ &  O2DiffDys & $\rightarrow$ &  HypGly \\ 
$p$-value: & \multicolumn{2}{c}{{ 0.039 (2018)}} & \multicolumn{2}{c}{{ 0.063 (2019)}} \\ 
\midrule[0.3pt]\bottomrule[1pt]
\end{tabular}
}
\end{small}
\end{center}
%%%%%\vspace{-0.15in}
%\vspace{-0.05in}
\end{table}

In Table~\ref{table:chain}, the chains possess a statistically strong relationship with patients in the Sepsis-3 cohort and correlate with clinical patterns that are often seen in sepsis. For example, Oxygen Diffusion Dysfunction (i.e., low oxygen saturation in the blood) is found to promote Renal Dysfunction and subsequent Oxygen Diffusion Dysfunction. Though not reflected in this table, septic patients could experience multiple chains simultaneously in addition to experiencing other discrete SADs simultaneous to events in a chain. This method to select and rank chains affords clinicians the ability to efficiently discover or follow those temporal patterns that differentiate septic patients from those experiencing organ injury caused by other diseases.

\subsubsection{Evaluating the quantitative performance} Due to the lack of time granularity of the time series data and the overly simple parametric form of the MHP model, we do not build a sequential prediction model to validate our method's usefulness. Instead, we quantitatively validate the usefulness of the identified GC chains by using them to construct features and apply a more sophisticated (but less interpretable) machine learning method to perform sequential prediction tasks. Here, we choose XGBoost due to its good performance (compared with other choices such as neural networks) in the sepsis prediction challenge \citep{physionet2019}.

\begin{table}[!htbp]
%\vspace{-0.08in}
\centering
\caption{Sepsis event prediction result. We can observe that incorporating the identified GC chains as input features (the proposed method) improves the accuracy compared with the benchmark vanilla XGBoost method.}\label{table:pred_boosting}
\vspace{0.1in}
\resizebox{.7\textwidth}{!}{%
\begin{tabular}{lccccc}
\toprule[1pt]\midrule[0.3pt]
 & \multicolumn{2}{c}{{In-sample (year 2018)}} & \multicolumn{2}{c}{Out-of-sample (year 2019)} \\ 
 \cmidrule(l){2-3}
\cmidrule(l){4-5}
 & Benchmark    &  Proposed & Benchmark    &  Proposed \\
Accuracy &  0.7183 &  0.7862 & 0.7214  & 0.7789 \\
Sensitivity &  0.7258 & 0.7983 & 0.7300 &  0.7930 \\
\midrule[0.3pt]
\bottomrule[1pt]
\end{tabular}
}
%\vspace{-0.08in}
\end{table}

In the benchmark XGBoost method, we use the mean values of the past 12 hours’ SADs as input features. In contrast, we additionally include binary variables indicating whether or not there exist chain patterns as shown in Table~\ref{table:chain} in the past 12 hours as the input features, to see whether or not this can improve the prediction accuracy and sensitivity. We use 5-fold cross validation (for grid search of hyperparameters in XGBoost) and train the model using 2018 data. We test the performance on 2019 data. The results are reported in Table~\ref{table:pred_boosting}, where we can observe improvements in both accuracy and sensitivity when predicting sepsis using our identified GC chains as input features. This suggests the usefulness of the identified GC chains; however, building a powerful prediction model with such GC chains is still on-going work.

%%%%\vspace{-0.05in}
\section{Conclusion}\label{sec:discussion}
%%%\vspace{-0.05in}

To conclude this paper, we briefly summarize the contribution and limitations of current work. We defer an extended discussion to Appendix~\ref{appendix:extended_discussion}. 
Our proposed method for Granger Causal chain discovery provides a novel and scalable approach to leverage clinical expertise to elucidate patterns of interest amongst large amounts of related EMR data. Though we do not build or validate a clinical alarm, this is a very useful and logical extension of this work. Additionally, knowledge from the GC chains could be used to estimate the risk of a future SAD (e.g., Renal Dysfunction) which might prompt a clinician to alter treatment (e.g., modify IV fluids therapy).
A limitation of this work stems from the grouped nature of many lab results and vital sign measurements. It is not uncommon for multiple patient features to be recorded in the EMR with identical timestamps which means that multiple SADs can occur simultaneously. This presents challenges to our point process model which can not capture relationships between simultaneously occurring SADs. This could be remedied by incorporating 
second or third-order interaction effect in ANOVA into the work to evaluate the effect of combined SADS on future patient states. 

Another limitation of the method arises from the 
way treatments are administered. Some treatments (i.e., antibiotics) are dosed on an interval whereas others (i.e., vasopressors) are dosed continuously. This results in a higher number of ``vasopressor'' events than antibiotic events for certain patients and can lead to the false conclusion that vasopressors are inhibiting antibiotics which is not an expected finding. Possible solutions include representing antibiotics as a continuous medication similar to vasopressors so that the continuous effects of antibiotics are appreciated by the model.

\section*{Acknowledgment}

The work of Song Wei and Yao Xie is partially supported by an NSF CAREER CCF-1650913, and NSF DMS-2134037, CMMI-2015787, CMMI-2112533, DMS-1938106, DMS-1830210, and an Emory Hospital grant.

\bibliographystyle{plainnat}  
\bibliography{ref} 

\newpage

\appendices

%\begin{KeepFromToc}
\addcontentsline{toc}{section}{Appendix} % Add the appendix text to the document TOC
\part{\centering \LARGE Appendix of Granger Causal Chain Discovery for Sepsis-Associated Derangements via Continuous-Time Hawkes Processes} % Start the appendix part

\topskip0pt
\vspace*{\fill}

\parttoc % Insert the appendix TOC
%\faketableofcontents
%\tableofcontents
%\end{KeepFromToc}

%\vfill
\vspace*{\fill}

\newpage

\section{Additional Details for the Proposed Method}

\subsection{Empirical challenge}\label{appendix:d3example}

In the $d=3$ illustrative example, the background intensities are $0.2,0.5,0.05$, $\beta = 0.6$ and the excitation/inhibition matrix $A$ is
$$
A = \begin{pmatrix}
0.1 & 0.2 & -0.3\\
-0.1 & 0.1 & 0 \\
0.5 & 0 & 0.5
\end{pmatrix}, \quad 
\tilde A = \begin{pmatrix}
0.1 & 0.2 & \alpha_{13}\\
\alpha_{21} & 0.1 & 0 \\
0.5 & 0 & 0.5
\end{pmatrix}.
$$
For illustrative purposes, we generate 200 sequences with $T = 120$ and evaluate the log likelihood on the ground truth background intensities and $\beta = 0.6$ as well as parameter $\tilde A$.
We plot the true log likelihood \eqref{eq:true_lik} as a function of $(\alpha_{13},\alpha_{21})$ in Figure~\ref{fig:opt_landscape}, where we observe that the likelihood becomes intractable around a very small neighborhood around the empirical optimizer.

\subsection{Algorithm}\label{appendix:add_alg}

\subsubsection{Phase 1: projected gradient descent}
Firstly, we perform PGD; instead of projecting back to the feasible region $\Theta$ \eqref{eq:feasible_region}, we simply set all negative entries to zeros (which is a subset of the feasible region). Thus, this phase is computationally friendly and can easily converge to a neighborhood of the empirical optimizer. Complete details of this phase can be found the in the following Algorithm~\ref{algo:2_phase_1}.

\begin{algorithm}[htp]
    \caption{Phase 1: Projected Gradient Descent.}\label{algo:2_phase_1}
    \begin{flushleft}
    \justifying
    \textbf{Input: } Data $(u_1, t_1),\dots,(u_N, t_N)$, learning rate $\gamma_1$, stopping criterion (e.g., a certain number of iterations), decay parameter $\beta_0$.
    \end{flushleft}
    
    \begin{flushleft}
    \justifying
    \textbf{Initialization: } Random initialize $\hat \mu_0$; Random or zero initialize $\hat A_0$; Iteration index $t=1$
    \end{flushleft}
    
    \begin{itemize}
        \item[] \textbf{while} stopping criterion NOT fulfilled \textbf{do}
    \begin{itemize}
        \item[\textbf{1}] Gradient Step: calculate the gradient
        \begin{align*}
        \nabla_{\mu} \tilde \ell &= \nabla_{\mu} \tilde \ell(\hat \mu_{t-1},\hat A_{t-1};\beta_0), \\
        \nabla_{A} \tilde \ell &= \nabla_{A} \tilde \ell(\hat \mu_{t-1},\hat A_{t-1};\beta_0),
        \end{align*}
        and then perform regular gradient descent with constant step length 
        \begin{align*}
            \hat \mu_t & \leftarrow \hat \mu_{t-1} + \gamma_1 \nabla_{\mu} \tilde \ell/\|\nabla_{\mu} \tilde \ell\|_2, \\
            \hat A_t & \leftarrow \hat A_{t-1} + \gamma_1 \nabla_{A} \tilde \ell/\|\nabla_{A} \tilde \ell \|_F
        \end{align*}
        \item[\textbf{2}] Projection Step: project all negative entries of $\hat \mu_t$ and $\hat A_t$ back to zeros
        $$\hat \mu_t \leftarrow (\hat \mu_t)^{+} , \quad  \hat A_t \leftarrow (\hat A_t)^{+}$$
        \item[\textbf{3}] Update the iteration index $t \leftarrow t+1$
    \end{itemize}
    \textbf{end while}
    \end{itemize}
    
\begin{flushleft}
    \textbf{Return: } $\hat \mu_t, \  \hat A_t$. 
    \end{flushleft}
\end{algorithm}

%\vspace{-.15in}

\newpage

\subsubsection{Phase 2: target index identification} After the initial phase, the iterate is in a neighborhood of the empirical optimizer, and we only need to further optimize those variables whose ground truth values are negative. To identify such variables, we use the gradient row vectors' $L_2$ norms as the indicator --- if the gradient's norm is large, the corresponding row vector in $A$ may not converge to its empirical optimizer, suggesting that there could be negative entries in this row. Complete details of this variable/node identification procedure can be found in Algorithm~\ref{algo:2_phase_2_index} below.

\begin{algorithm}[htp]
    \caption{Phase 2: Identification of the target indices for Batch Coordinate Gradient Descent.}\label{algo:2_phase_2_index}
    \begin{flushleft}
    \justifying
    \textbf{Input: } Data $(u_1, t_1),\dots,(u_N, t_N)$, decay parameter $\beta_0$, a percentage threshold $p \in (0,1)$ and the output of Phase 1 $(\hat \mu_0, \ \hat A_0)$ in Algorithm~\ref{algo:2_phase_1}.
    \end{flushleft}
    
    \begin{itemize}
        \item[\textbf{1}] Sort the index set $I=\{1,\dots,d\}$ such that the corresponding vector $\tilde L_2$ norm of the rows of the gradient matrix is in a descending order:
        \begin{align*}
            J &=\{J_1,\dots,J_d\} \\
            &= \arg{\rm sort} \{\|\nabla_{A_j} \tilde \ell(\hat \mu_0, \hat A_0;\beta_0)\|_2: \  j \in I \},
        \end{align*}
        where $A_j$ denotes the $j$-th row of matrix $A$
        
        \item[\textbf{2}] Find the minimum number $\tilde{d}$ of the first few indices in $J$ such that the corresponding gradient row norms make up $p \times 100 \%$ of the total $F$-norm of matrix $A$, $\|\nabla_{A} \tilde \ell(\hat \mu_0, \hat A_0;\beta_0)\|_F$:
        \begin{align*}
            \quad \quad  \tilde{d} & =  {\rm argmin} \Bigg\{K: \\
            & \quad \quad \sum_{k=1}^K \|\nabla_{A_{J_{k}}} \tilde \ell(\hat \mu_0, \hat A_0;\beta_0)\|_2^2 \geq p \|\nabla_{A} \tilde \ell(\hat \mu_0, \hat A_0;\beta_0)\|_F^2 \Bigg\}
        \end{align*}
    \end{itemize}
    
\begin{flushleft}
    \textbf{Return: } $J, \ \tilde{d}$. 
    \end{flushleft}
\end{algorithm}

\begin{algorithm*}[!htp]
    \caption{Phase 2: (Batch) Coordinate Descent for Selected variables.}\label{algo:2_phase_2_main}
\begin{flushleft}
    \textbf{Input: } Data $(u_1, t_1),\dots,(u_N, t_N)$, learning rate $\gamma_2$, decay parameter $\beta_0$, a learning rate threshold $\gamma_0$, target index set $J, \ \tilde{d}$ from Algorithm~\ref{algo:2_phase_2_index} and the output of Phase 1 in Algorithm~\ref{algo:2_phase_1} $(\hat \mu_0, \ \hat A_0)$.
\end{flushleft}    
    
    % \textbf{Initialization: } Initialize at the output of Phase 1 in Algorithm~\ref{algo:2_phase_1} $\hat \mu_0, \ \hat A_0$.
    
    \begin{itemize}
        
        \item[] \textbf{for} $i = J_1, \dots, J_{\tilde{d}} $ \textbf{do}
        \begin{itemize}
        \item[] Initialize iteration index $t=1$ and learning rate $\tilde{\gamma_2} = {\gamma_2}$
             \item[] \textbf{while} $\tilde{\gamma_2} > \gamma_0$ \textbf{do}
    \begin{itemize}
        \item[\textbf{1}] Calculate the gradient
        $$\nabla_{A_i} \tilde \ell = \nabla_{A_i} \tilde \ell(\hat \mu_{0},\hat A_{t-1};\beta_0),$$
        and then perform regular gradient descent with constant step length 
        $$ (\hat A_{t})_{i} \leftarrow  (\hat A_{t-1})_{i} + \tilde{\gamma_2} \nabla_{A_i} \tilde \ell/\|\nabla_{A_i} \tilde \ell \|_2$$
        \item[\textbf{2}] \textbf{if} $\|\nabla_{A_i} \tilde \ell\|_2 < \|\nabla_{A_i} \tilde \ell(\hat \mu_{0},\hat A_{t};\beta_0)\|_2$ \textbf{then}
        
        \hspace{.25in} Decrease learning rate $\tilde{\gamma_2} \leftarrow \tilde{\gamma_2}/2$ and cancel the update $(\hat A_{t})_{i} \leftarrow  (\hat A_{t-1})_{i}$
        
        \textbf{end if}

        \item[\textbf{3}] Update the iteration index $t \leftarrow t+1$
    \end{itemize}
    \textbf{end while}
    
            \item[] Update $(\hat A_0)_i \leftarrow  (\hat A_t)_{i}$ and initialize iteration index $t=1$ and learning rate $\tilde{\gamma_2} = {\gamma_2}$
             \item[] \textbf{while} $\tilde{\gamma_2} > \gamma_0$ \textbf{and} $(\hat \mu_{t-1})_{i} > 0$ \textbf{do}
    \begin{itemize}
        \item[\textbf{1}] Gradient Step: Calculate the gradient
        $$\nabla_{\mu_i} \tilde \ell = \nabla_{\mu_i} \tilde \ell(\hat \mu_{t},\hat A_{0};\beta_0),$$
        and then perform regular gradient descent with constant step length 
        $$ (\hat \mu_{t})_{i} \leftarrow  (\hat \mu_{t-1})_{i} + \tilde{\gamma_2} \nabla_{\mu_i} \tilde \ell/\|\nabla_{\mu_i} \tilde \ell \|_2$$
        \item[\textbf{2}] Projection Step: $(\hat \mu_{t})_{i} \leftarrow \left((\hat \mu_{t})_{i}\right)^+$
        
    \item[\textbf{3}] \textbf{if} $\|\nabla_{\mu_i} \tilde \ell\|_2 < \|\nabla_{\mu_i} \tilde \ell(\hat \mu_{t},\hat A_{0};\beta_0)\|_2$ \textbf{then}
        
        \hspace{.25in} Decrease learning rate $\tilde{\gamma_2} \leftarrow \tilde{\gamma_2}/2$ and cancel the update $(\hat \mu_{t})_{i} \leftarrow  (\hat \mu_{t-1})_{i}$
        
        \textbf{end if}
        
        \item[\textbf{4}] Update the iteration index $t \leftarrow t+1$
    \end{itemize}
    \textbf{end while}
    
    \item[] Update $(\hat \mu_0)_i \leftarrow  (\hat \mu_t)_{i}$
        \end{itemize}

    \textbf{end for}
    \end{itemize}
    
\begin{flushleft}
    \textbf{Return: } $\hat \mu_0, \  \hat A_0$. 
\end{flushleft}
\end{algorithm*}

%\vspace{-.15in}

\subsubsection{Phase 2: batch coordinate descent}
\begin{figure}[!htp]
%%%%\vspace{-0.1in}
\centerline{
\includegraphics[width = \textwidth]{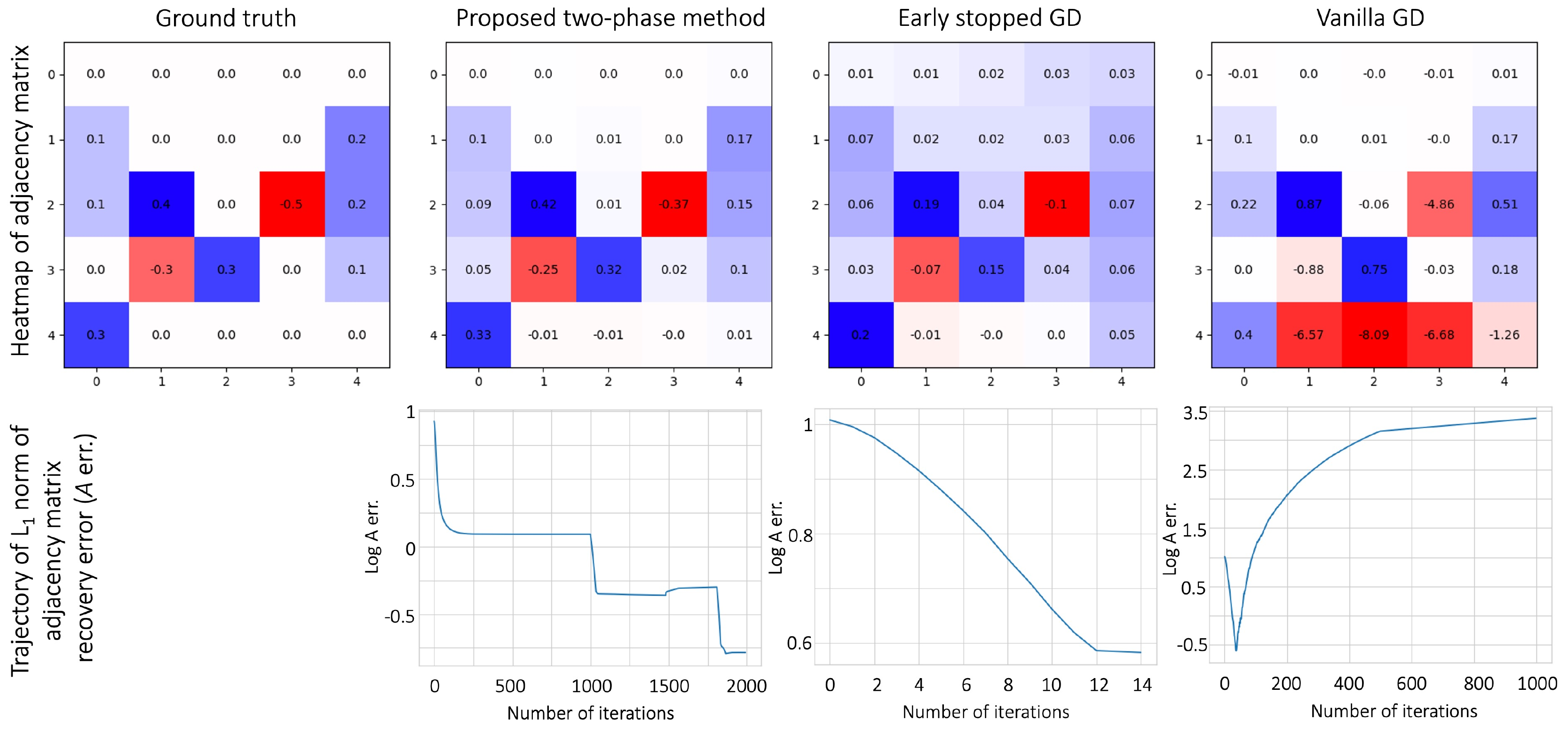}}
%%%%\vspace{-0.25in}
\caption{Comparison to GD on a $d=5$ toy example. We can see the early stopped GD only converge to a large neighborhood of the optimizer whereas the vanilla GD with fixed learning rate can easily diverge due to the existence of negative entries in the adjacency matrix. Nevertheless, early stopped GD still cannot output accurate results; see our numerical simulation in Section~\ref{subsec:exp3_comparison_benchmark}.}
\label{fig:compare_GD_illus}
%%%%\vspace{-0.3in}
\end{figure}

Now, we are ready to perform batch coordinate descent for those selected variables. In this phase, we do not apply any projection to allow for possible negative entries in the iterate. However, similar to vanilla GD (as evidenced in Figure~\ref{fig:compare_GD_illus}), the instability/divergence issue still exists. To handle this issue, we design a learning rate decaying scheme based on the gradient-norm.

\begin{figure}[!htp]
%%%%\vspace{-0.1in}
\centerline{
\includegraphics[width = .95\textwidth]{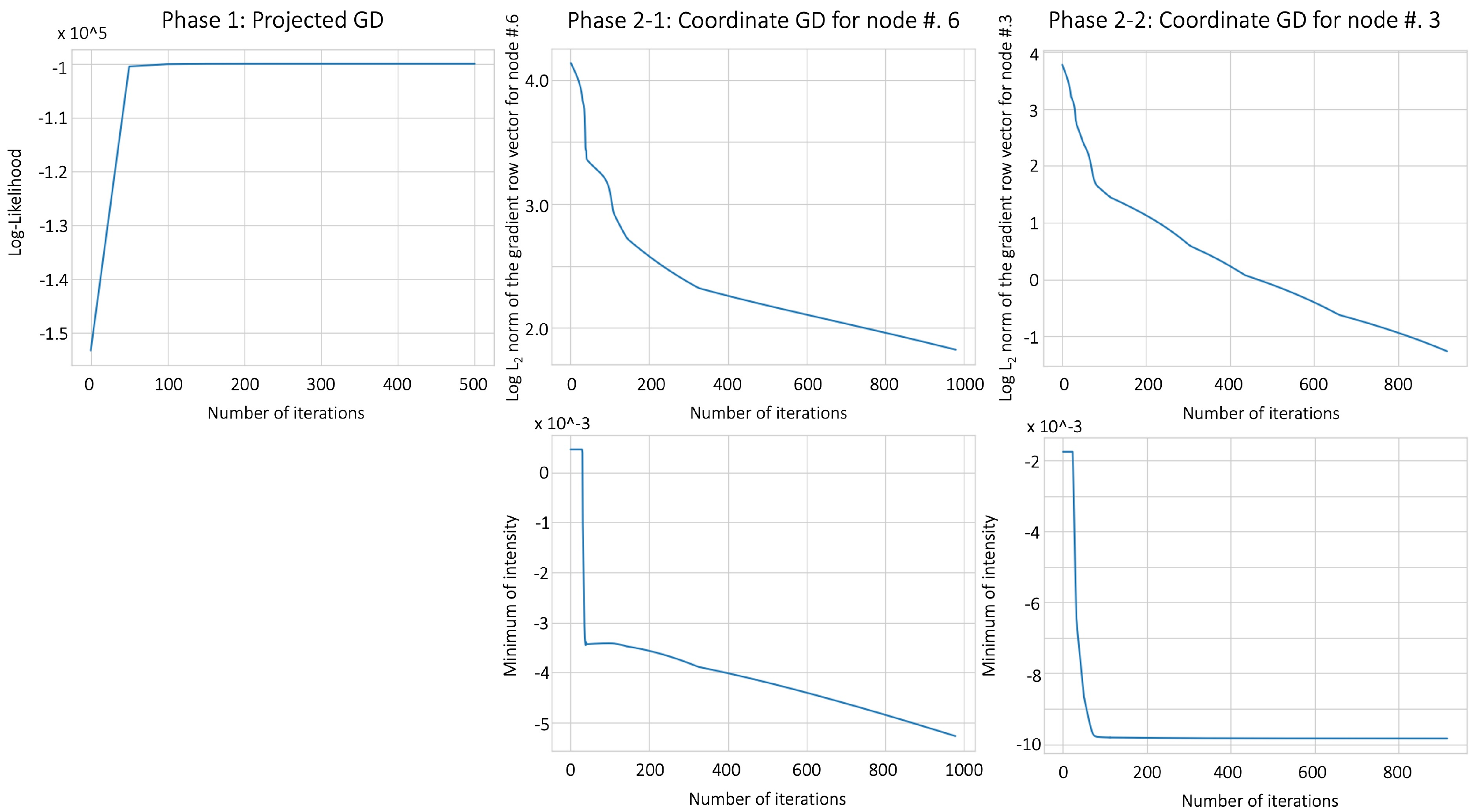}}
%%%%\vspace{-0.25in}
\caption{Evidence of convergence for the $d=10$ toy example in Section~\ref{sec:method}. In phase 1, the log likelihood gradually increases. In phase 2, the log likelihood becomes intractable as the minimum intensity (at even occurrence times) turns negative. Nevertheless, the decreasing gradient-norm shows it can serve as a criterion for the algorithm's convergence.}
\label{fig:illus_criterion}
%%%%\vspace{-0.3in}
\end{figure}
Since the minimum intensity becomes negative in phase 2, making the log likelihood no longer well-defined (as shown in Figure~\ref{fig:illus_criterion}), we propose to use gradient $L_2$ norm to determine whether or not we should decrease the learning rate in Phase 2. The intuition is straightforward:
as we can see in Figure~\ref{fig:illus_criterion}, due to the convexity of the objective function, we know the gradient should approach zero as we approach the global optimizer. We show the complete detail of the second part of phase 2 in Algorithm~\ref{algo:2_phase_2_main}.

\section{Additional Details for the Numerical Simulation}

\subsection{Experimental details}\label{appendix:exp_detail}
\subsubsection{Synthetic data generation}\label{appendix:syn_data_generation}
We use Lewis' thinning algorithm \citep{lewis1979simulation} to simulate the MHP data. In this experiment, we fix the decay parameter $\beta = 0.8$ and randomly initialize the background intensity vector $\mu \in \mathbb{R}^d$ and the network adjacency matrix $A \in \mathbb{R}^{d \times d}$. Each entry of $\mu$ (and $A$) is uniformly random number in $[0,0.1]$ (and $[0,0.4]$). Next, we apply vanilla GD to minimize function $h({A})=\operatorname{tr}\left(e^{{A}}\right)-d$ to zero, to obtain a directed acyclic graph \citep{zheng2018dags}. Last but not least, we randomly assign random negative entries ($U([-0.5,0])$) to $A$ and round all entries of $\mu$ and $A$ to one decimal place. Even though the synthetic data only considers DAG as input, we want to mention that we do not explicitly use the structure in the learning process. Most importantly, the $L_1$ estimation error is also reported, which can validate the effectiveness of our method for most types of random graphs.

\subsubsection{Evaluation metrics}
Here, we report the following metrics
\begin{itemize}
    \item  absolute value of $\beta$ estimation error ($\beta$ err.);
    \item  vector $L_1$ norm of $\mu$ estimation error ($\mu$ err.);
    \item  vector $L_1$ norm of $A$ estimation error ($A$ err.);
    \item  Hamming Distance between ground truth and estimated (after thresholding) adjacency matrix $A$;
    \item  Structural Hamming Distance between ground truth and estimated (after thresholding) adjacency matrix $A$.
\end{itemize}
Here, the metric $A$ err. aims to evaluate how the estimated values deviate from the ground truth whereas metrics HD and SHD aim to show how well the estimated graph structure/pattern resembles the ground truth. The Hamming Distance is defined to be the number of differences between the supports of ground truth and estimated (after thresholding) adjacency matrix $A$.
For SHD, it is defined as the total number of edge additions, deletions, and reversals needed to convert the estimated DAG into the true DAG. Its evaluation is readily implemented in a python package \texttt{cdt} \citep{kalainathan2019causal} and we refer readers to Appendix D.2 in \citet{zheng2018dags} for a more detailed definition.

Last but not least, the hard thresholding of estimated adjacency matrix is done by zeroing out all entries whose absolute values are smaller than a threshold. We start with the smallest value of all absolute values of the non-zero entries in the estimated matrix as the threshold. We gradually increase this threshold until the resulting graph is a DAG (i.e., $h(A) = 0$).

\subsection{Additional results}\label{appendix:add_exp}

\subsubsection{Additional results for experiment 1}
Here, we report the error metrics for the known $\beta$ case in higher dimension ($d = 20$) for completeness in Table~\ref{table:exp:consistency_d20}.
However, due to the computational limitation, we only report the results for a smaller number of sequences and shorter time horizons. Nevertheless, we can still numerically verify the consistency.

\begin{table}[!htp]
%%%%%\vspace{-0.1in}
\caption{Performance of proposed method when $\beta$ is assumed to be known for $d=20$ case. We can still observe that all error metrics are decreasing with either an increasing number of sequences or time horizon, which further verifies our finding on the empirical consistency. }\label{table:exp:consistency_d20}
%%%%%\vspace{-0.25in}
\begin{center}
\begin{small}
\resizebox{.75\textwidth}{!}{%
\begin{tabular}{lcccc}
\multicolumn{5}{c}{{Varying sequence time horizon $T$ (sequence number fixed to be 1)}} \\ 
\toprule[1pt]

T & 500 & 2000 & 5000 & 10000 \\% Column names row
\cmidrule(l){2-5}
 % In-table horizontal line
\texttt{$\mu$ err.$^\star$} & 58.45  (19.23) & 33.75  (11.3) & 24.88  (8.87) & 21.01  (7.67)   \\
\texttt{$A$ err.} & 39.43 (15.93) & 12.07 (4.14) & 6.98 (2.26) & 5.51 (2.19)\\
\texttt{$A$ HD} & 0.24 (0.064) & 0.068 (0.051) & 0.032 (0.037) & 0.025 (0.044)  \\
\texttt{$A$ SHD} & 96.0 (25.91) & 27.5 (20.66) & 13.0 (14.97) & 10.0 (17.64)  \\
\bottomrule[1pt]
%\multicolumn{5}{l}{$\star$ we omit $\times 10^{-2}$ in the value due to space consideration.}
\\
\end{tabular}
}
\end{small}
\end{center}

%%%%%\vspace{-0.1in}
%\caption{Performance of our method with different  for $d=20$ case. We can observe that the error metrics are decreasing with increasing sequence number.}\label{table:exp:T_consistency_d20_2}
%%%%%\vspace{-0.25in}

\begin{center}
\begin{small}
\resizebox{.58\textwidth}{!}{%
\begin{tabular}{lccc}
\multicolumn{4}{c}{{Varying sequence number (time horizon $T = 500$)}} \\
\toprule[1pt]
%\midrule[0.3pt]
Seq. Num. & 1 & 10 & 20 \\% Column names row
\cmidrule(l){2-4}
 % In-table horizontal line
\texttt{$\mu$ err.$^\star$} & 57.91  (19.71) & 24.16  (8.84) & 20.34  (7.56)   \\
\texttt{$A$ err.} & 39.62 (16.33) & 6.88 (2.42) & 5.24 (2.22)   \\
\texttt{$A$ HD} & 0.24 (0.068) & 0.03 (0.034) & 0.02 (0.041)  \\
\texttt{$A$ SHD} & 96.0 (27.47) & 12.0 (13.8) & 8.0 (16.52)  \\
\bottomrule[1pt]
\multicolumn{4}{l}{$\star$ we omit $\times 10^{-2}$ in the value due to space consideration.}
\\
\end{tabular}
}
\end{small}
\end{center}
%%%%%\vspace{-0.1in}
\end{table}

\begin{table*}[!htp]
%\centering
%%%%\vspace{-0.25in}
\caption{Median and standard deviation of (1) $L_1$ norm of $\mu$ estimation error ($\mu$ err.) ; (2) $L_1$ norm of $A$ estimation error ($A$ err.); (3) Hamming Distance between ground truth and estimated adjacency matrix $A$ (\texttt{$A$ HD}); (4) Structural Hamming Distance between ground truth and estimated adjacency matrix $A$ (\texttt{$A$ SHD}). The last row corresponds to selected $\beta$ by using matrix $F$-norm of phase-two $A$ gradient as the criterion, where we can observe its performance is not as good as what we have when we use end-of-phase 1 log likelihood as the criterion in Table~\ref{table:exp:beta_grid_saerch_lik}.}\label{table:exp:beta_grid_saerch_grad}
%\vspace{-0.1in}
\begin{center}
\begin{small}
\resizebox{\textwidth}{!}{%
\begin{tabular}{lcccccccccccccr}
\toprule[1pt]\midrule[0.3pt]
& \multicolumn{4}{c}{$d=5$} & \multicolumn{4}{c}{$d=10$}  \\
$\beta$ & \texttt{$\mu $ err.$^\star$} & \texttt{$A$ err.} & \texttt{$A$ HD} & \texttt{$A$ SHD} & \texttt{$\mu $ err.$^\star$} & \texttt{$A$ err.} & \texttt{$A$ HD} & \texttt{$A$ SHD}  \\
\cmidrule(l){2-5}
\cmidrule(l){6-9}
        0.4 & 4.14   (2.79) & 1.34 (0.25) & 0.08 (0.093) & 2.0 (2.33) & 8.6   (3.83) & 4.37 (0.64) & 0.04 (0.054) & 4.0 (5.47) \\ 
        0.5 & 4.27   (2.93) & 1.11 (0.25) & 0.08 (0.087) & 2.0 (2.19) & 10.49   (4.4) & 3.49 (0.61) & 0.02 (0.035) & 2.0 (3.52) \\ 
        0.6 & 4.26   (2.88) & 0.86 (0.24) & 0.04 (0.074) & 1.0 (1.86) & 9.84   (4.34) & 2.58 (0.63) & 0.02 (0.025) & 2.0 (2.51)\\ 
        0.7 & 3.78   (2.74) & 0.63 (0.22) & 0.02 (0.062) & 0.5 (1.55) & 8.68   (4.02) & 1.87 (0.63) & 0.01 (0.026) & 1.0 (2.61)\\ 
        0.8 & 3.57   (2.6) & 0.54 (0.23) & 0.0 (0.051) & 0.0 (1.29) & 7.33   (3.72) & 1.4 (0.49) & 0.01 (0.018) & 1.0 (1.89)\\ 
        0.9 & 3.53   (2.52) & 0.64 (0.22) & 0.0 (0.038) & 0.0 (0.96) & 6.46   (3.52) & 1.53 (0.42) & 0.01 (0.021) & 1.0 (2.17)\\ 
        1 & 3.79   (2.45) & 0.91 (0.24) & 0.0 (0.035) & 0.0 (0.88) & 7.09   (3.45) & 1.84 (0.39) & 0.01 (0.022) & 1.0 (2.2)\\ 
        1.1 & 3.68   (2.35) & 1.14 (0.28) & 0.0 (0.024) & 0.0 (0.61) & 7.61   (3.4) & 2.3 (0.34) & 0.02 (0.027) & 2.0 (2.71)\\ 
        1.2 & 3.85   (2.23) & 1.31 (0.36) & 0.0 (0.029) & 0.0 (0.73) & 8.05   (3.37) & 2.71 (0.38) & 0.03 (0.034) & 3.0 (3.47)\\ 
        $-$ & 4.04   (2.46) & 0.98 (0.53) & 0.0 (0.028) & 0.0 (0.7) & 7.47   (3.26) & 2.34 (0.66) & 0.02 (0.03) & 2.0 (3.0)\\ 
\midrule[0.3pt]
\bottomrule[1pt]
\multicolumn{8}{l}{$\star$ we omit $\times 10^{-2}$ in the value due to space consideration.}
\\
\end{tabular}
}
\end{small}
\end{center}
%\vspace{-0.1in}
\end{table*}

\subsubsection{Additional results for experiment 2}
Since the log likelihood is intractable in phase 2 coordinate descent, we also try the $F$-norm of the gradient w.r.t. $A$ as the GoF metric to select the hyperparameter $\beta$. To be precise, we will select the $\beta$ with the smallest gradient (w.r.t. $A$) $F$-norm as the estimated decay parameter. We report the results in  %both Figure~\ref{fig:exp_beta_grid} and
Table~\ref{table:exp:beta_grid_saerch_grad}, where we can see it does not perform well for $d=10$ case, compared to using end-of-phase 1 log likelihood as the GoF metric in Table~\ref{table:exp:beta_grid_saerch_lik}.
Last but not least, since we have end-of-phase 1 log likelihood as a valid GoF metric, we can apply $L_1$ penalty in the MLE formulation \eqref{eq:MLE} to impose sparse structure on graph structure $A$.

\subsection{Comparison with benchmarks}

\subsubsection{Benchmarks}\label{appendix:benchmark}
Here, we give details on the benchmark methods that we compare with in our numerical simulation. For vanilla GD, we use a fixed learning rate without any decaying scheme and for early stopped GD, we decrease the learning rate by half whenever the gradient-norm increases (as the log likelihood could be intractable). For the re-start time method, we only give a graphical illustration here and refer readers to \citet{bonnet2021maximum,bonnet2022inference} for more details.

\begin{figure}[!htp]
\centerline{
\includegraphics[width = .4\textwidth]{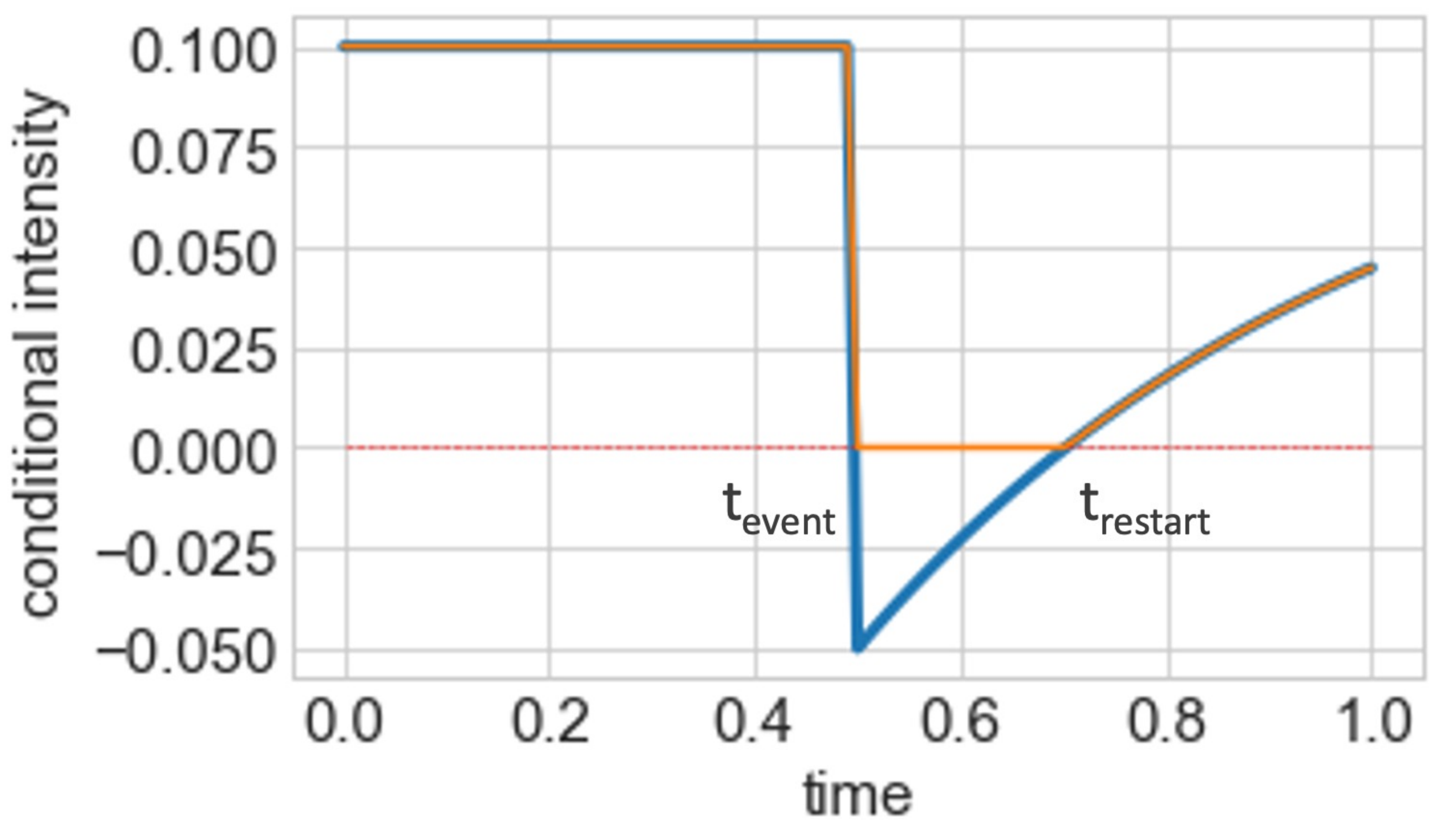}}
\caption{Illustration of the re-start time \citep{bonnet2021maximum} for $d=1$ case. The blue line is the surrogate intensity whereas the orange line is the true intensity. 
}
\label{fig:restart}
%\vspace{-0.05in}
\end{figure}

As one can see in Figure~\ref{fig:restart}, the orange line corresponds to the true conditional intensity $\lambda_i(t)$ \eqref{eq:condi_intensity_def} whereas the blue line corresponds to the surrogate $\tilde \lambda_i(t)$ \eqref{eq:surrogate_intensity}. To evaluate the log likelihood, for each event occurrence time $t_{\rm event}$ we need to calculate a re-start time $t_{\rm restart}$ as shown in the above figure. Clearly, when we are faced with a multivariate problem, we need to calculate the re-start time for ALL variables at each event occurrence time. Thus, this calculation is very computationally intensive and we did encounter this issue when performing our numerical simulation in $d = 20$ case. Moreover, the authors did not talk about how to generalize the re-start time method to multi-sequence settings and therefore we apply SGD to handle the multi-sequence case.

\subsubsection{Additional results for experiment 3}\label{appendix:runtime}

In Section~\ref{subsec:exp3_comparison_benchmark}, we show that our proposed approach is able to output a much more accurate estimate of the Granger causal graph compared with benchmark approaches. Here, we demonstrate another major advantage of our proposed approach, which is the scalablility. Here, we report the average run time (in seconds) of the gradient calculation for both re-start and our proposed approaches in Table~\ref{table:exp:runtime}, where we can observe that our proposed approach is much faster compared with the re-start time method. 
Moreover, we also include the run time analysis of the overall convergence, since it is challenging to obtain the theoretical convergence analysis for both approaches. We only consider $d = 5, T = 240$ case here to demonstrate the benefit: for the re-start time approach, it costs 608.94 (12.89) seconds to converge whereas it only takes 444.71 (12.98) seconds for our proposed method to converge. Note that we do not consider larger dimensions (nor longer sequences) since it takes too long for the re-start time approach to converge; we do not consider EM algorithm either since (i) it needs modification to fix the intractability issue under our setting and (ii) it introduces too many parameters as $T$ becomes larger (and so does $N$).

\begin{table*}[!htp]
%\centering
%%%%\vspace{-0.25in}
\caption{Run time analysis (unit: second) of the gradient evaluation for our proposed approach and re-start time approach. Here, we report the mean and standard deviation over 100 independent trials. We can observe that the gradient evaluation for our proposed approach is much faster than that of the re-start time approach, which severs as a numerical evidence of our complexity analysis in Table~\ref{tab:complexity}.}\label{table:exp:runtime}
%\vspace{-0.1in}
\begin{center}
\begin{small}
\resizebox{\textwidth}{!}{%
\begin{tabular}{lcccccc}
\toprule[1pt]\midrule[0.3pt]
& \multicolumn{3}{c}{$d=5$} & \multicolumn{3}{c}{$d=10$}  \\
$T$ & 240 & 480 & 960 & 240 & 480 & 960 \\
\cmidrule(l){2-4}
\cmidrule(l){5-7}
Re-start & 0.0053 (0.0005)	& 0.0107 (0.0005) & 0.0205 (0.0010) & 0.0299 (0.0017) & 0.0595 (0.0023) & 0.1178 (0.0021) \\
Proposed & 0.0015 (0.0002)	& 0.0030 (0.0001) & 0.0057 (0.0002) & 0.0156 (0.0006) & 0.0316 (0.0009) & 0.0643 (0.0009) \\
\midrule[0.3pt]
\bottomrule[1pt]
\end{tabular}
}
\end{small}
\end{center}
%\vspace{-0.1in}
\end{table*}

The above complexity and run time analysis indicate that our approach has the potential to scale to larger graphs and longer observation sequences. However, it should be noted that the primary contribution of our work is not only this scalability --- an efficient and reliable learning approach for linear multivariate Hawkes processes with inhibiting effects is currently missing in the literature, and this research direction is demonstrated to be meaningful through a real-data example presented in Section~\ref{sec:real_data_exp}. The contribution of our methodology to this direction, as well as the findings from applying our proposed approach to the real-data example, are equally significant. Since our real-world motivating example only consists of 21 nodes, it suffices to demonstrate the effectiveness as well as improved accuracy of our proposed approach using moderate-size graphs compared with baseline approaches.

\section{Additional Details for the Real Data Example}

\subsection{Data description and SAD construction}\label{appendix:lab_vital_SAD}

All patient data for each encounter was binned into hourly windows that began with hospital admission and ended with discharge. If more than one measurement occurred in an hour, then the mean of the values was recorded. To ensure that model training was performed on data series of similar lengths, a 24-hour subset of the full patient encounter was selected for analysis. For patients in the Sepsis-3 cohort, the 24-hour window ended when the patient met the Sepsis-3 criteria (SEP3 time). For the Non-Septic cohort, the window was centered (i.e.,12 hours before and after) upon the first time a patient had a SOFA score $\geq$ 2 (SOFA time). In some instances the event of interest (i.e., SOFA time or SEP3 time) occurred close to admission or discharge, resulting in a truncated data set of $<$24 hours.

\begin{table*}[!htp]
%%%%%\vspace{-0.2in}
\caption{SAD construction based on thresholding observed vital signs and Labs via medical knowledge. In our study, we incorporate in total 37 patient features, including 31 Labs and 6 vital signs.}\label{table:SADcutoff}
%\vspace{-0.1in}
\begin{center}
\begin{small}
\resizebox{\textwidth}{!}{%
\begin{tabular}{lllc}
\toprule[1pt]\midrule[0.3pt]
Full name (Abbrev.) &  Type & Measurement name & Abnormal threshold  \\ % Column names row
\midrule[0.3pt]
 % In-table horizontal line
\textbf{Renal Dysfunction} (RenDys) & Lab & creatinine & $>1.3$ \\
& & blood\_urea\_nitrogen\_(bun) & $>20$ \\
\cmidrule(l){3-4}
\textbf{Electrolyte Imbalance} (LyteImbal) & Lab & calcium & $>10.5$ \\
& & chloride & $<98$ OR $>106$ \\
& & magnesium & $<1.6$ \\
& & potassium & $>5.0$ \\
& & phosphorus & $>4.5$ \\
\cmidrule(l){3-4}
\textbf{Oxygen Transport Deficiency} (O2TxpDef) & Lab & hemoglobin & $<12$ \\
\cmidrule(l){3-4}
\textbf{Coagulopathy} (Coag) & Lab & partial\_prothrombin\_time\_(ptt) & $>35$ \\
& & fibrinogen & $<233$ \\
& & platelets & $<150000$ \\
& & d\_dimer & $>0.5$ \\
& & thrombin\_time & $>20$ \\
& & prothrombin\_time\_(pt) & $>13$ \\
& & inr & $>1.5$ \\
\cmidrule(l){3-4}
\textbf{Malnutrition} (MalNut) & Lab & transferrin & $<0.16$ \\
& & prealbumin & $<16$ \\
& & albumin & $<3.3$ \\
\cmidrule(l){3-4}
\textbf{Cholestatsis} (Chole) & Lab & bilirubin\_direct & $>0.3$ \\
& & bilirubin\_total & $>1.0$ \\
\cmidrule(l){3-4}
\textbf{Hepatocellular Injury} (HepatoDys) & Lab & aspartate\_aminotransferase\_(ast) & $>40$ \\
& & alanine\_aminotransferase\_(alt) & $>40$ \\
& & ammonia & $>70$ \\
\cmidrule(l){3-4}
\textbf{Acidosis} (Acidosis) & Lab & base\_excess & $<-3$ \\
& & ph & $<7.32$ \\
\cmidrule(l){3-4}
\textbf{Leukocyte Dysfunction} (LeukDys) & Lab & white\_blood\_cell\_count & $<4$ OR $>12$ \\
\cmidrule(l){3-4}
\textbf{Hypercarbia} (HypCarb) & Lab & end\_tidal\_co2 & $>45$ \\
& & partial\_pressure\_of\_carbon\_dioxide\_(paco2) & $>45$ \\
\cmidrule(l){3-4}
\textbf{Hyperglycemia} (HypGly) & Lab & glucose & $>125$ \\
\cmidrule(l){3-4}
\textbf{Mycardial Ischemia} (MyoIsch) & Lab & troponin & $>0.04$ \\
\cmidrule(l){3-4}
% \textbf{Vasopressor Support} (VasoSprt) & Treatment & norepinephrine\_dose\_weight & $>0$ \\
% & & epinephrine\_dose\_weight & $>0$ \\
% & & dobutamine\_dose\_weight & $>0$ \\
% & & dopamine\_dose\_weight & $>0$ \\
% & & phenylephrine\_dose\_weight & $>0$ \\
% & & vasopressin\_dose\_weight & $>0$ \\
% \cmidrule(l){3-4}
\textbf{Tissue Ischemia} (TissueIsch) & Lab & base\_excess & $<-3$ \\
& & lactic\_acid & $>2.0$ \\
\cmidrule(l){3-4}
\textbf{Diminished Cardiac Output} (DCO) & Vital signs & best\_map & $<65$ \\
\cmidrule(l){3-4}
\textbf{CNS Dysfunction} (CNSDys) & Vital signs & gcs\_total\_score & $<14$ \\
\cmidrule(l){3-4}
\textbf{Oxygen Diffusion Dysfunction} (O2DiffDys) & Vital signs & spo2 & $<92$ \\
& & fio2 & $>21$ \\
\cmidrule(l){3-4}
\textbf{Thermoregulation Dysfunction} (ThermoDys) & Vital signs & temperature & $<36$ OR $>38$ \\
\cmidrule(l){3-4}
\textbf{Tachycardia} (Tachy) & Vital signs & pulse & $>90$ \\
\midrule[0.3pt]\bottomrule[1pt]
\end{tabular}
}
\end{small}
\end{center}
%%%%%\vspace{-0.15in}
%%%%%\vspace{-0.25in}
\end{table*}

The raw features from the EMR data we are using include:
(i) {Vital Signs}
--- in ICU environments these are normally recorded at hourly intervals, however, patients on the floor may only have vital signs measured once every 8 hours. 
(ii) {Lab Results}
--- These tests are most commonly collected once every 24 hours; however, this collection frequency may change based on the severity of a patient's illness. In this study, we include in total 6 vital signs and 31 Labs. Those vital signs and Labs are presented in the SAD construction table in Table~\ref{table:SADcutoff}. As those measurement names explain themselves, we do not give further descriptions of the meaning of those measurements.

% \begin{figure*}[htp]
% %%%%%\vspace{-0.05in}
% \centerline{
% \includegraphics[width = 0.55\textwidth]{tex files/DAG_Hawkes/realdata/mat_sep1_num.pdf}
% %\includegraphics[width = .5\textwidth]{tex files/DAG_Hawkes/realdata/mat_full_num.pdf}
% }
% %%%\vspace{-0.15in}
% \caption{Estimated adjacency matrix for Sepsis-3 cohort.}
% \label{fig:mat_num_sep}

% %\vspace{0.2in}

% \centerline{
% %\includegraphics[width = .5\textwidth]{tex files/DAG_Hawkes/realdata/mat_sep_num.pdf}
% \includegraphics[width = .55\textwidth]{tex files/DAG_Hawkes/realdata/mat_full_num.pdf}}
% %%%\vspace{-0.15in}
% \caption{Estimated adjacency matrix for full patient cohort.}
% \label{fig:mat_num_full}
% %%%%%\vspace{-0.15in}
% \end{figure*}

\subsection{GC graph recovery}\label{appendix:realdata_GC}

In our real data experiment, we additionally add $L_1$ regularization to enforce sparse structure on the estimated adjacency matrix, i.e., the estimate is obtained by solving the following optimization problem:
\begin{equation}\label{eq:MLE+L1}
    \hat \mu, \hat A = {\rm argmin}_{\mu,A \in \Theta} - \tilde \ell(\mu,A;\beta) + \lambda_1 \|A\|_1,
\end{equation}
where $\tilde \ell(\mu,A;\beta)$ is the surrogate log likelihood defined in \eqref{eq:surrogate_lik}. Here, both regularization parameter $\lambda_1$ and decaying rate parameter $\beta$ are hyperparameters. We use grid search to find their optimal values. To be precise, in our real data experiments, we search $\lambda_1 \in \{0.05,0.1,0.15,0.2,0.25,0.3,0.35,0.4,$ $0.5\}$ and $\beta \in \{0.01,0.015,0.02,$ $0.025,0.03,0.035,0.04\}$. 

\begin{figure}[!htp]
%\vspace{-0.1in}
\centerline{
\includegraphics[width = .85\textwidth]{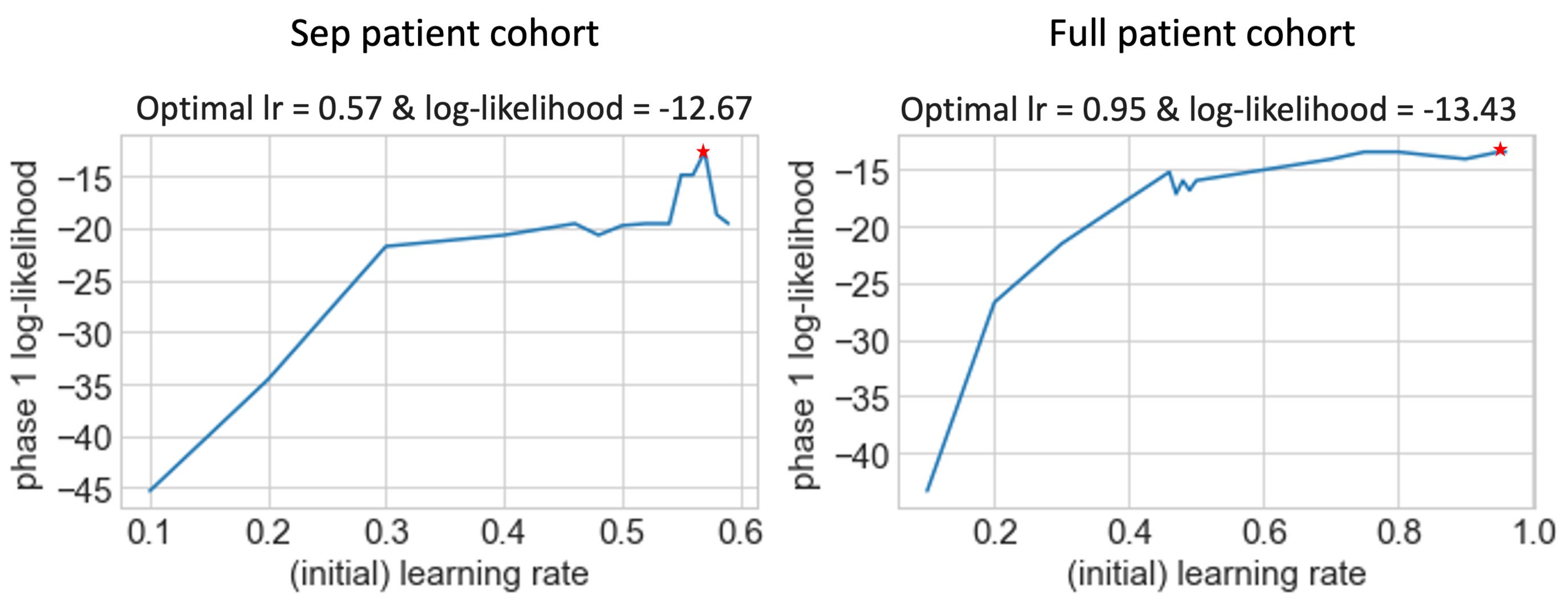}}
%\vspace{-0.13in}
\caption{End-of-phase 1 log likelihood with different learning rates. Each point on the curve corresponds to the optimal selected $(\lambda_1, \beta)$ pair on our selected grid based on the end-of-phase 1 log likelihood. We do not try larger $\gamma_1$ on the full patient cohort due to divergence. The optimal learning rate is marked with a red star and its value is on the top of the corresponding figure.}
\label{fig:lr_selection}
%\vspace{-0.1in}
\end{figure}

Most importantly, we also find that our method is very sensitive to the phase 1 learning rate $\gamma_1$ choice. Thus, we also use end-of-phase 1 likelihood as the GoF criterion to select the best $\gamma_1$. In phase 1, we run 1000 iterations and reduce the learning rate by half every 200 iterations. The trade-off is that choosing an overly small learning rate cannot guarantee convergence within the limited iterations whereas an overly large learning rate will lead to divergence or over-fitting.
Here, when we train the model for Sepsis-3 cohort, we search $\gamma_1 \in \{0.6,0.59,0.58,0.57,0.56,0.55,0.54,0.53,0.52,0.5,0.48,0.46,0.4,$ $0.3,0.2,0.1\}$. For the full patient model, we search $\gamma_1 \in \{0.97,0.96,$ $0.95,0.9,0.85,0.8,0.75,0.7,0.5,0.49,0.48,0.47,$ $0.46,0.4,$ 
$0.3,0.2,0.1\}$.

\begin{figure*}[htp]
%%%%%\vspace{-0.05in}
\centerline{
\includegraphics[width = .65\textwidth]{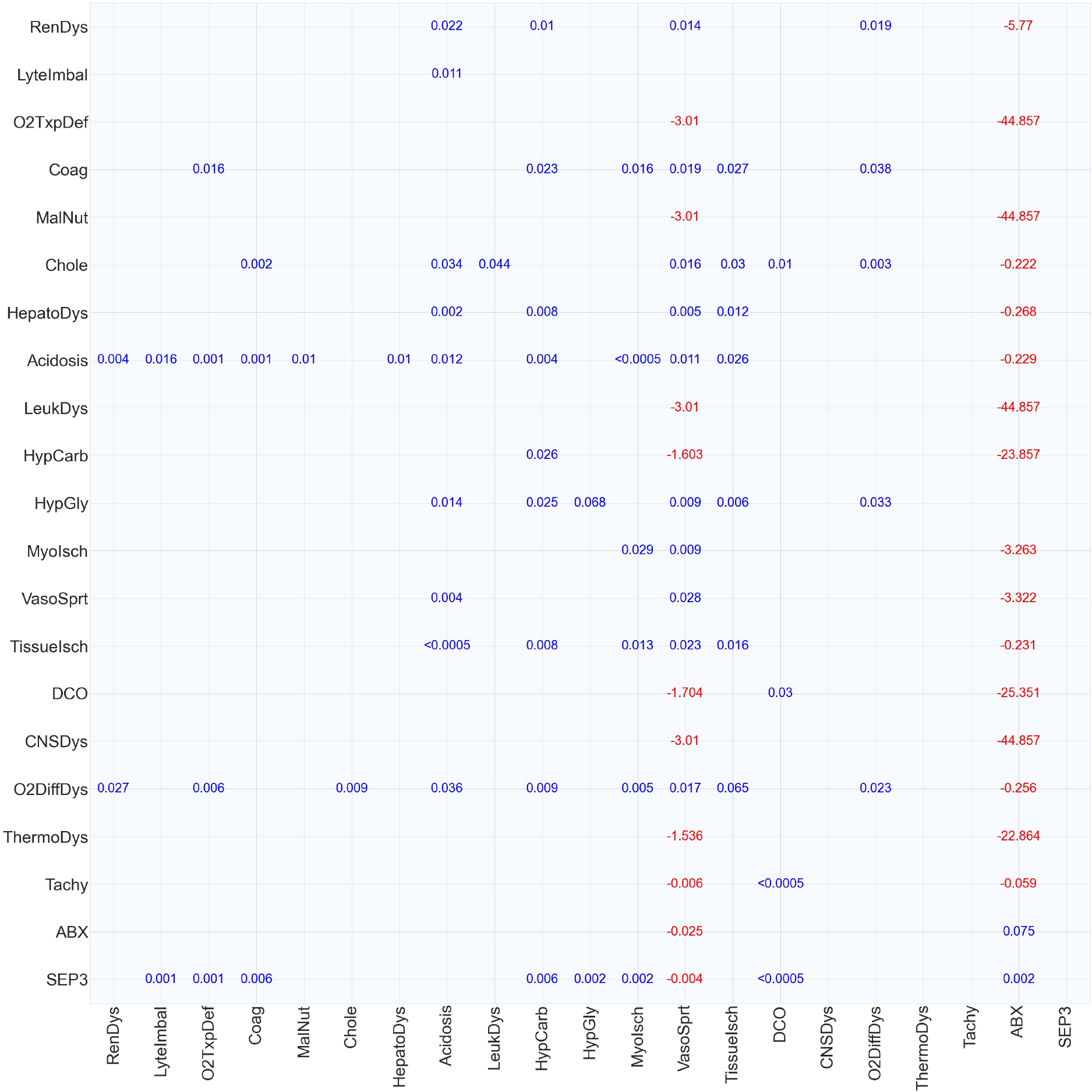}
}

%\vspace{0.1in}

\centerline{
\includegraphics[width = .65\textwidth]{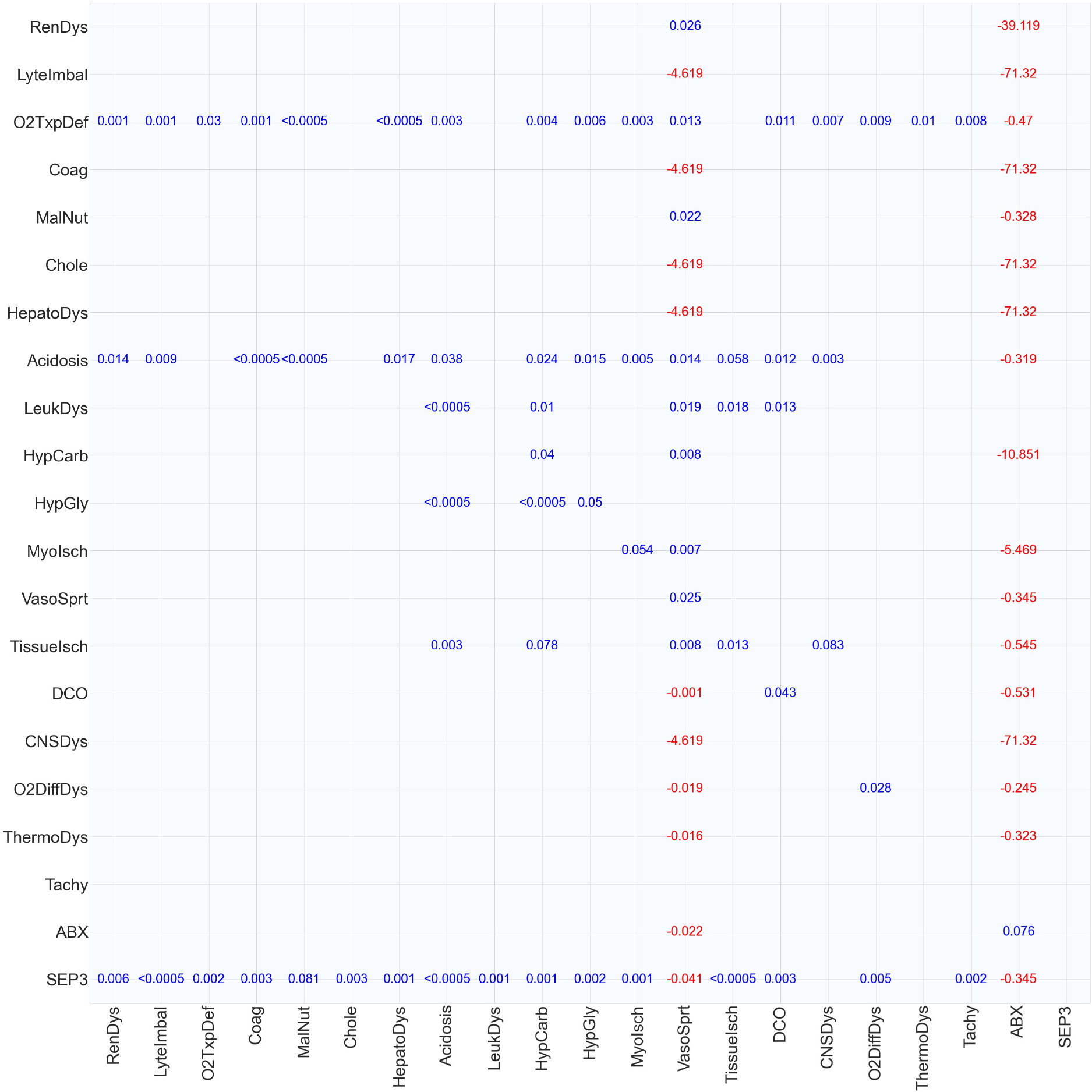}
}
%%%\vspace{-0.15in}
\caption{Estimated adjacency matrices for Sepsis-3 cohort (top) and full patient cohort (bottom).}
\label{fig:mat_num}
\end{figure*}

We fit the model using the data from the year 2018 and report the end-of-phase 1 log likelihood versus learning rate in Figure~\ref{fig:lr_selection}, where each point on the curve corresponds to the optimal selected $(\lambda_1, \beta)$ pair based on the end-of-phase 1 log likelihood. For the Sepsis-3 model, the optimal hyperparameters are $\gamma_1 = 0.57, \lambda_1 = 0.25, \beta = 0.03$; for the full patient model, the optimal hyperparameters are: $\gamma_1 = 0.95, \lambda_1 = 0.5, \beta = 0.02$. We also want to mention that for $\gamma_1 \geq 0.97$, our method diverges for the full patient cohort and therefore we do not try larger $\gamma_1$ choices.

The background intensities for both models are: (i) For the Sepsis-3 model, we have estimated background intensities are $.0326$,$.0424$, $.0698$, and $.0363$ for Electrolyte Imbalance, Thermoregulation Dysfunction, Tachycardia, and Sepsis, respectively, whereas the rest of the SAEs' background intensities are all zeros; (ii) For full patient model, we have $.0676, .0514, .0719$ for Hyperglycemia, Thermoregulation Dysfunction, Tachycardia, respectively, and the rest are all zeros. Most importantly, the estimated mutual excitation and inhibition matrices are reported in
Figure~\ref{fig:mat_num}, respectively.

\subsection{GC chain discovery and identification}\label{appendix:realdata_chain}

Now, we are ready to identify the chain structures/patterns which are unique in Sepsis-3 cohort. 
However, those entries could correspond to chains with different orders and lengths.
Therefore, we still need to use a more convincing method to validate the uniqueness of those possible chains in the Sepsis-3 GC graph (and not in the full patient GC graph). 

\subsubsection{GC chain identification via Fisher's exact test}
Let us take a two-node graph as an example: for RenDys and O2DiffDys, we can extract a 2-by-2 sub-adjacency matrix in Sepsis-3 GC graph. As we can see, only the self-exciting effect is zero in this sub-matrix, leading to 2 possible length-2 chain structures: RenDys $\rightarrow$ O2DiffDys and O2DiffDys $\rightarrow$ RenDys, 3 possible length-3 chain structures: RenDys $\rightarrow$ O2DiffDys $\rightarrow$ O2DiffDys, O2DiffDys $\rightarrow$ O2DiffDys $\rightarrow$ RenDys and O2DiffDys $\rightarrow$ RenDys $\rightarrow$ O2DiffDys, and so on.

\begin{table}[H]
\caption{Data for Fisher's exact test}\label{table:fisher}
%\vspace{-.15in}
\begin{center}
    \begin{tabular}{r|cc}
 count & Sepsis-3  & Non-sepatic   \\ \hline
 with chain&  a  & b \\
 without chain&  c & d
\end{tabular}
\end{center}
\end{table}

To identify which chain structure leads to the entries in the GC graph, we perform Fisher's exact test, which is a powerful hypothesis testing tool on the equality of ratios between two groups when the sample size is relatively small. To be precise, given a chain structure, assume we observe Table~\ref{table:fisher},
the estimates of probabilities of such chain structure's occurrence in both groups are given by 
$$\hat p_{\rm sep} = \frac{a}{a+c}, \quad \hat p_{\rm non-sep} = \frac{b}{b+d}.$$
Under the null hypothesis $$H_0: p_{\rm sep} = p_{\rm non-sep},$$ the probability of observing the above Table~\ref{table:fisher} is
\begin{align*}
    p&=\frac{\left(\begin{array}{c}
a+b \\
a
\end{array}\right)\left(\begin{array}{c}
c+d \\
c
\end{array}\right)}{\left(\begin{array}{c}
n \\
a+c
\end{array}\right)}=\frac{\left(\begin{array}{c}
a+b \\
b
\end{array}\right)\left(\begin{array}{c}
c+d \\
d
\end{array}\right)}{\left(\begin{array}{c}
n \\
b+d
\end{array}\right)}\\
&=\frac{(a+b) !(c+d) !(a+c) !(b+d) !}{a ! b ! c ! d ! n !},
\end{align*}
where $n=a+b+c+d$. This is the $p$-value of Fisher's exact test and we will reject $H_0$ if this value is smaller than a chosen significance level $\alpha$ (e.g., 0.05 or 0.1).

\subsubsection{GC chain discovery}
In our real data experiment, for a given chain structure, we first extract data with all events in the chain to form Table~\ref{table:fisher} and then calculate a $p$-value based on the above formula. We choose $\alpha \approx 0.1$ as the significance level. 
We say a discovered chain (by our GC graphs) is identified to be (significantly) unique in the Sepsis-3 cohort if the corresponding $p$-value of Fisher's exact test is smaller than our selected significance level $\alpha$.

In the following, we will report the data table (as shown in Table~\ref{table:fisher}), ratio estimates $\hat p_{\rm sep}, \hat p_{\rm non-sep}$ and the $p$-value for all significantly unique chains. We report the chains which are significantly unique in the Sepsis-3 cohort in both year 2018 (i.e., in-sample test) and year 2019 (i.e., out-of-sample test) in Table~\ref{table:chain1}. In addition, we also report the chains which are only significantly unique in the Sepsis-3 cohort in one year in Tables~\ref{table:chain2} (the year 2018) and \ref{table:chain3} (the year 2019).

\begin{table}[!htp]
%%%%\vspace{-0.2in}
\caption{GC chains which are significantly unique in Sepsis-3 cohort in both 2018 and 2019.}\label{table:chain1}
%\vspace{-.1in}
\begin{center}
\begin{small}
\resizebox{.55\textwidth}{!}{%
\begin{tabular}{crcccc}
\toprule[1pt]
\toprule[1pt]
Chain: & TissueIsch  & $\rightarrow$ &  O2DiffDys \\ 
\cmidrule(l){1-6}
& & \multicolumn{2}{c}{{2018}} & \multicolumn{2}{c}{{2019}} \\ 
& & Sep & Non-Sep & Sep & Non-sep \\ 
\cmidrule(l){2-6}
& Count & 17 & 20 & 19 & 35 \\ 
&  & 5 & 30 & 6 & 28 \\ 
& Ratio & 0.772 & 0.4 & 0.76 & 0.555 \\ 
& $p$-value & \multicolumn{2}{c}{{ 0.004 }} & \multicolumn{2}{c}{{ 0.092 }} \\ 
\toprule[1pt]
Chain: & O2DiffDys  & $\rightarrow$ &  RenDys & $\rightarrow$ &  O2DiffDys \\ 
\cmidrule(l){1-6}
& & \multicolumn{2}{c}{{2018}} & \multicolumn{2}{c}{{2019}} \\ 
& & Sep & Non-Sep & Sep & Non-sep \\ 
\cmidrule(l){2-6}
& Count & 35 & 31 & 49 & 33 \\ 
&  & 14 & 25 & 17 & 34 \\ 
& Ratio & 0.714 & 0.553 & 0.742 & 0.492 \\ 
& $p$-value & \multicolumn{2}{c}{{ 0.107 }} & \multicolumn{2}{c}{{ 0.004 }} \\ 
\toprule[1pt]
Chain: & VasoSprt  & $\rightarrow$ &  TissueIsch & $\rightarrow$ &  HepatoDys \\ 
\cmidrule(l){1-6}
& & \multicolumn{2}{c}{{2018}} & \multicolumn{2}{c}{{2019}} \\ 
& & Sep & Non-Sep & Sep & Non-sep \\ 
\cmidrule(l){2-6}
& Count & 3 & 3 & 3 & 2 \\ 
&  & 1 & 14 & 3 & 15 \\ 
& Ratio & 0.75 & 0.176 & 0.5 & 0.117 \\ 
& $p$-value & \multicolumn{2}{c}{{ 0.052 }} & \multicolumn{2}{c}{{ 0.088 }} \\ 
\toprule[1pt]
Chain: & LyteImbal  & $\rightarrow$ &  Acidosis & $\rightarrow$ &  O2DiffDys \\ 
\cmidrule(l){1-6}
& & \multicolumn{2}{c}{{2018}} & \multicolumn{2}{c}{{2019}} \\ 
& & Sep & Non-Sep & Sep & Non-sep \\ 
\cmidrule(l){2-6}
& Count & 8 & 7 & 9 & 13 \\ 
&  & 2 & 17 & 3 & 19 \\ 
& Ratio & 0.8 & 0.291 & 0.75 & 0.406 \\ 
& $p$-value & \multicolumn{2}{c}{{ 0.009 }} & \multicolumn{2}{c}{{ 0.088 }} \\ 
\toprule[1pt]
Chain: & Acidosis  & $\rightarrow$ &  O2DiffDys & $\rightarrow$ &  HypGly \\ 
\cmidrule(l){1-6}
& & \multicolumn{2}{c}{{2018}} & \multicolumn{2}{c}{{2019}} \\ 
& & Sep & Non-Sep & Sep & Non-sep \\ 
\cmidrule(l){2-6}
& Count & 6 & 13 & 9 & 13 \\ 
&  & 1 & 20 & 6 & 30 \\ 
& Ratio & 0.857 & 0.393 & 0.6 & 0.302 \\ 
& $p$-value & \multicolumn{2}{c}{{ 0.039 }} & \multicolumn{2}{c}{{ 0.063 }} \\ 
\bottomrule[1pt]
\bottomrule[1pt]
\end{tabular}
}

\end{small}
\end{center}
%\vspace{-0.1in}
\end{table}

\begin{table}[!htp]
\caption{GC chains which are only significantly unique in Sepsis-3 cohort in the year 2018.}\label{table:chain2}
\begin{center}
\begin{small}
%\vspace{-.1in}
\resizebox{0.68\textwidth}{!}{%
\begin{tabular}{crcccccc}
\toprule[1pt]
\toprule[1pt]
Chain: & MalNut  & $\rightarrow$ &  Acidosis \\ 
\cmidrule(l){1-8}
& & &\multicolumn{2}{c}{{2018}} & \multicolumn{2}{c}{{2019}} \\ 
& & & Sep & Non-Sep & Sep & Non-sep \\ 
\cmidrule(l){2-8}
& Count&  & 24 & 20 & 21 & 35 \\ 
&  & &  9 & 27 & 11 & 34 \\ 
& Ratio&  & 0.727 & 0.425 & 0.656 & 0.507 \\ 
& $p$-value&  & \multicolumn{2}{c}{{ 0.011 }} & \multicolumn{2}{c}{{ 0.198 }} \\ 
\toprule[1pt]
Chain: & RenDys  & $\rightarrow$ &  O2DiffDys \\ 
\cmidrule(l){1-8}
& & &\multicolumn{2}{c}{{2018}} & \multicolumn{2}{c}{{2019}} \\ 
& & & Sep & Non-Sep & Sep & Non-sep \\ 
\cmidrule(l){2-8}
& Count&  & 44 & 39 & 57 & 50 \\ 
&  & &  5 & 17 & 9 & 17 \\ 
& Ratio&  & 0.897 & 0.696 & 0.863 & 0.746 \\ 
& $p$-value&  & \multicolumn{2}{c}{{ 0.015 }} & \multicolumn{2}{c}{{ 0.125 }} \\ 
\toprule[1pt]
Chain: & TissueIsch  & $\rightarrow$ &  O2DiffDys & $\rightarrow$ &  HypGly \\ 
\cmidrule(l){1-8}
& & &\multicolumn{2}{c}{{2018}} & \multicolumn{2}{c}{{2019}} \\ 
& & & Sep & Non-Sep & Sep & Non-sep \\ 
\cmidrule(l){2-8}
& Count&  & 10 & 11 & 11 & 23 \\ 
&  & &  3 & 28 & 6 & 31 \\ 
& Ratio&  & 0.769 & 0.282 & 0.647 & 0.425 \\ 
& $p$-value&  & \multicolumn{2}{c}{{ 0.003 }} & \multicolumn{2}{c}{{ 0.164 }} \\ 
\toprule[1pt]
Chain: & Acidosis  & $\rightarrow$ &  RenDys & $\rightarrow$ &  O2DiffDys & $\rightarrow$ &  HypGly \\ 
\cmidrule(l){1-8}
& & &\multicolumn{2}{c}{{2018}} & \multicolumn{2}{c}{{2019}} \\ 
& & & Sep & Non-Sep & Sep & Non-sep \\ 
\cmidrule(l){2-8}
& Count&  & 5 & 3 & 4 & 2 \\ 
&  & &  1 & 8 & 4 & 11 \\ 
& Ratio&  & 0.833 & 0.272 & 0.5 & 0.153 \\ 
& $p$-value&  & \multicolumn{2}{c}{{ 0.049 }} & \multicolumn{2}{c}{{ 0.146 }} \\ 
\toprule[1pt]
Chain: & TissueIsch  & $\rightarrow$ &  O2DiffDys & $\rightarrow$ &  RenDys & $\rightarrow$ &  O2DiffDys \\ 
\cmidrule(l){1-8}
& & &\multicolumn{2}{c}{{2018}} & \multicolumn{2}{c}{{2019}} \\ 
& & & Sep & Non-Sep & Sep & Non-sep \\ 
\cmidrule(l){2-8}
& Count&  & 7 & 2 & 6 & 7 \\ 
&  & &  5 & 19 & 4 & 17 \\ 
& Ratio&  & 0.583 & 0.095 & 0.6 & 0.291 \\ 
& $p$-value&  & \multicolumn{2}{c}{{ 0.004 }} & \multicolumn{2}{c}{{ 0.129 }} \\ 
\toprule[1pt]
Chain: & Acidosis  & $\rightarrow$ &  LyteImbal & $\rightarrow$ &  Acidosis & $\rightarrow$ &  O2DiffDys \\ 
\cmidrule(l){1-8}
& & &\multicolumn{2}{c}{{2018}} & \multicolumn{2}{c}{{2019}} \\ 
& & & Sep & Non-Sep & Sep & Non-sep \\ 
\cmidrule(l){2-8}
& Count&  & 5 & 4 & 6 & 7 \\ 
&  & &  5 & 20 & 6 & 25 \\ 
& Ratio&  & 0.5 & 0.166 & 0.5 & 0.218 \\ 
& $p$-value&  & \multicolumn{2}{c}{{ 0.084 }} & \multicolumn{2}{c}{{ 0.134 }} \\ 
\toprule[1pt]
Chain: & LyteImbal  & $\rightarrow$ &  Acidosis & $\rightarrow$ &  O2DiffDys & $\rightarrow$ &  HypGly \\ 
\cmidrule(l){1-8}
& & &\multicolumn{2}{c}{{2018}} & \multicolumn{2}{c}{{2019}} \\ 
& & & Sep & Non-Sep & Sep & Non-sep \\ 
\cmidrule(l){2-8}
& Count&  & 4 & 4 & 6 & 7 \\ 
&  & &  2 & 17 & 5 & 22 \\ 
& Ratio&  & 0.666 & 0.19 & 0.545 & 0.241 \\ 
& $p$-value&  & \multicolumn{2}{c}{{ 0.044 }} & \multicolumn{2}{c}{{ 0.127 }} \\ 
\bottomrule[1pt]
\bottomrule[1pt]
\end{tabular}
}

\end{small}
\end{center}
%\vspace{-0.1in}
%%%%%\vspace{-0.25in}
\end{table}

\begin{table*}[!htp]
\caption{GC chains which are NOT significantly unique in Sepsis-3 cohort 2018, but significantly unique in 2019.}\label{table:chain3}

%%%%\vspace{-0.3in}

\begin{center}
\begin{small}

\resizebox{0.6\textwidth}{!}{%
\begin{tabular}{crcccccc}
\toprule[1pt]
\toprule[1pt]
Chain: & O2DiffDys  & $\rightarrow$ &  RenDys \\ 
\cmidrule(l){1-8}
& & &\multicolumn{2}{c}{{2018}} & \multicolumn{2}{c}{{2019}} \\ 
& & & Sep & Non-Sep & Sep & Non-sep \\ 
\cmidrule(l){2-8}
& Count&  & 40 & 43 & 55 & 47 \\ 
&  & &  9 & 13 & 11 & 20 \\ 
& Ratio&  & 0.816 & 0.767 & 0.833 & 0.701 \\ 
& $p$-value&  & \multicolumn{2}{c}{{ 0.633 }} & \multicolumn{2}{c}{{ 0.1 }} \\ 
\toprule[1pt]
Chain: & VasoSprt  & $\rightarrow$ &  Acidosis & $\rightarrow$ &  RenDys \\ 
\cmidrule(l){1-8}
& & &\multicolumn{2}{c}{{2018}} & \multicolumn{2}{c}{{2019}} \\ 
& & & Sep & Non-Sep & Sep & Non-sep \\ 
\cmidrule(l){2-8}
& Count&  & 2 & 7 & 4 & 3 \\ 
&  & &  1 & 9 & 2 & 13 \\ 
& Ratio&  & 0.666 & 0.437 & 0.666 & 0.187 \\ 
& $p$-value&  & \multicolumn{2}{c}{{ 0.582 }} & \multicolumn{2}{c}{{ 0.053 }} \\ 
\toprule[1pt]
Chain: & RenDys  & $\rightarrow$ &  O2DiffDys & $\rightarrow$ &  RenDys \\ 
\cmidrule(l){1-8}
& & &\multicolumn{2}{c}{{2018}} & \multicolumn{2}{c}{{2019}} \\ 
& & & Sep & Non-Sep & Sep & Non-sep \\ 
\cmidrule(l){2-8}
& Count&  & 32 & 31 & 47 & 35 \\ 
&  & &  17 & 25 & 19 & 32 \\ 
& Ratio&  & 0.653 & 0.553 & 0.712 & 0.522 \\ 
& $p$-value&  & \multicolumn{2}{c}{{ 0.324 }} & \multicolumn{2}{c}{{ 0.032 }} \\ 
\toprule[1pt]
Chain: & VasoSprt  & $\rightarrow$ &  TissueIsch & $\rightarrow$ &  Acidosis & $\rightarrow$ &  RenDys \\ 
\cmidrule(l){1-8}
& & &\multicolumn{2}{c}{{2018}} & \multicolumn{2}{c}{{2019}} \\ 
& & & Sep & Non-Sep & Sep & Non-sep \\ 
\cmidrule(l){2-8}
& Count&  & 2 & 7 & 3 & 0 \\ 
&  & &  1 & 7 & 3 & 15 \\ 
& Ratio&  & 0.666 & 0.5 & 0.5 & 0.0 \\ 
& $p$-value&  & \multicolumn{2}{c}{{ 1.0 }} & \multicolumn{2}{c}{{ 0.015 }} \\ 
\toprule[1pt]
Chain: & O2DiffDys  & $\rightarrow$ &  RenDys & $\rightarrow$ &  O2DiffDys & $\rightarrow$ &  RenDys \\ 
\cmidrule(l){1-8}
& & &\multicolumn{2}{c}{{2018}} & \multicolumn{2}{c}{{2019}} \\ 
& & & Sep & Non-Sep & Sep & Non-sep \\ 
\cmidrule(l){2-8}
& Count&  & 32 & 31 & 48 & 31 \\ 
&  & &  17 & 25 & 18 & 36 \\ 
& Ratio&  & 0.653 & 0.553 & 0.727 & 0.462 \\ 
& $p$-value&  & \multicolumn{2}{c}{{ 0.324 }} & \multicolumn{2}{c}{{ 0.002 }} \\ 
\toprule[1pt]
Chain: & RenDys  & $\rightarrow$ &  O2DiffDys & $\rightarrow$ &  RenDys & $\rightarrow$ &  O2DiffDys \\ 
\cmidrule(l){1-8}
& & &\multicolumn{2}{c}{{2018}} & \multicolumn{2}{c}{{2019}} \\ 
& & & Sep & Non-Sep & Sep & Non-sep \\ 
\cmidrule(l){2-8}
& Count&  & 31 & 31 & 46 & 34 \\ 
&  & &  18 & 25 & 20 & 33 \\ 
& Ratio&  & 0.632 & 0.553 & 0.696 & 0.507 \\ 
& $p$-value&  & \multicolumn{2}{c}{{ 0.433 }} & \multicolumn{2}{c}{{ 0.033 }} \\ 
\toprule[1pt]
Chain: & O2DiffDys  & $\rightarrow$ &  RenDys & $\rightarrow$ &  O2DiffDys & $\rightarrow$ &  Coag \\ 
\cmidrule(l){1-8}
& & &\multicolumn{2}{c}{{2018}} & \multicolumn{2}{c}{{2019}} \\ 
& & & Sep & Non-Sep & Sep & Non-sep \\ 
\cmidrule(l){2-8}
& Count&  & 28 & 26 & 40 & 25 \\ 
&  & &  17 & 20 & 20 & 33 \\ 
& Ratio&  & 0.622 & 0.565 & 0.666 & 0.431 \\ 
& $p$-value&  & \multicolumn{2}{c}{{ 0.67 }} & \multicolumn{2}{c}{{ 0.015 }} \\ 
\toprule[1pt]
Chain: & LyteImbal  & $\rightarrow$ &  Acidosis & $\rightarrow$ &  O2DiffDys & $\rightarrow$ &  Coag \\ 
\cmidrule(l){1-8}
& & &\multicolumn{2}{c}{{2018}} & \multicolumn{2}{c}{{2019}} \\ 
& & & Sep & Non-Sep & Sep & Non-sep \\ 
\cmidrule(l){2-8}
& Count&  & 4 & 4 & 6 & 7 \\ 
&  & &  5 & 18 & 5 & 23 \\ 
& Ratio&  & 0.444 & 0.181 & 0.545 & 0.233 \\ 
& $p$-value&  & \multicolumn{2}{c}{{ 0.184 }} & \multicolumn{2}{c}{{ 0.072 }} \\ 
\bottomrule[1pt]
\bottomrule[1pt]
\end{tabular}
}

\end{small}
\end{center}
%%%%%\vspace{-0.15in}
%%%%%\vspace{-0.25in}
\end{table*}

\section{Extended Discussion}\label{appendix:extended_discussion}

Here, in addition to the potential improvements in Section~\ref{sec:discussion}, we point out several other directions that are worthy of exploration in the future. 

From a modeling perspective, an obvious drawback of Granger Causality is pointed out by \citet{eichler2010granger}, namely that $A$ Granger-causes $B$ does not imply that intervening in $A$ would affect (the distribution on) $B$. A famous example is that the purchase of Christmas trees Granger-causes Christmas but this is clearly not the case.  
Another example is that disease $B$ will lead to symptom $A$, say fever. Typically a person will first observe $A$ and after diagnostic testing be diagnosed with $B$. Therefore, we should expect $A$ Granger-caused $B$; however, we should not expect that taking fever-reducing medication (e.g., Tylenol), will help cure disease $B$ (though it will alter the value of $A$).
There are two possible ways to handle this issue: First, we should NOT treat the observed time as the exact event occurrence time. Instead, the exact occurrence time could be earlier than this observed time. By incorporating this time uncertainty, there are chances that we can infer the true causal structure. Second, we can try to conduct counterfactual analysis by sampling counterfactual trajectories. This might be a direct application of \citet{noorbakhsh2021counterfactual}, though we may need to extend this work to a multivariate setting.

Additionally, we want to remark on the connection between Granger causality and Structural Causal Model (SCM)-based causality:
Efforts have been made in the discrete-time time series setting to extend the recovery of Granger causality (GC) via vector autoregressive (VAR) models to the recovery of causal directed acyclic graphs (DAGs) via Structural VAR (SVAR) models. The main distinction between the two lies in the consideration of instantaneous effects among the nodes in the graph, where this implicit order ``at the same time index'' may represent the causal structure. One challenge within this framework is the lack of identifiability guarantee, as there could be multiple causal DAGs that correspond to the same data. Most importantly, it remains unclear how to extend this framework to the continuous-time point process setting, as there are no instantaneous effects when we observe the exact event occurrence times. In this work, we focus on Granger causality and we will leave the causality from the SCM-based approaches for future discussion.

Our method takes a simple linear form since it is suggested by the domain expert that by properly extracting the feature and the patient cohort, simple models can perform quite well. This suggests that our method is suitable for moderate-size graphs where knowledge can be distilled and the dimensionality can be reduced based on expert opinions (clustering can also be used as an alternative when expert knowledge is missing). In addition, based on the advice from the domain expert, it would be better to consider patients with similar sofa score and this suggests our model may not be able to hand patients with heterogeneous SADs' interactions.
This over-simplified linear MHP model may explain why it fails to perform the sequential prediction task: on one hand, the dynamics within the human body are too complex to be captured by a simple linear relationship; on the other hand, the exponential decaying kernel is problematic since an abnormal lab test result or vital sign for a small time period is ``acceptable'' whereas prolonged abnormality/dysfunction could lead to severe problem. Therefore, properly choosing the decaying kernel is very important. Nevertheless, since (1) it is easy to develop a scalable method to fit a simple model and (2) we do demonstrate its usefulness in identifying unique chain patterns in a real data experiment, we believe this is still a meaningful contribution to literature. Future work on how to utilize such chain patterns to perform sequential prediction, or how to use more complex yet explainable models (e.g., \citet{wei2021inferringb}) to fulfill such prediction purpose, needs to be done.

Lastly, from the optimization perspective, we only give a heuristic, i.e., our proposed two-phase algorithm, that works well in practice. Reformulating the problem via Lagrangian duality and using projected gradient descent to solve the dual problem could both work well in practice and enjoy a strong convergence guarantee in theory. To be precise, the expansive projection for primal variables onto feasible region $\Theta$ \eqref{eq:feasible_region} is reduced to the simple projection for dual variables onto $\mathbb{R}_+^{d \times d}$. Moreover, it is also worthwhile (numerically) to explore the effect of other structures/regularizations on adjacency matrix, e.g., DAG structure \citep{zheng2018dags,ng2020role} or low-rank structure \citep{fang2020low}.

\end{document}